\documentclass[aps,epsf,preprint,showpacs]{revtex4}
\usepackage{amsmath}
\usepackage{epsfig}

\begin{document}

\title{Dynamic phase transition of the Blume-Capel model in an oscillating magnetic field}

\author{Erol Vatansever$^1$}

\author{Nikolaos~G. Fytas$^2$}

\affiliation{$^1$Department of Physics, Dokuz Eyl\"{u}l
University, TR-35160, Izmir-Turkey}

\affiliation{$^2$Applied Mathematics Research Centre, Coventry
University, Coventry  CV1 5FB, United Kingdom}

\date{\today}

\begin{abstract}
We employ numerical simulations and finite-size scaling techniques
to investigate the properties of the dynamic phase transition that
is encountered in the Blume-Capel model subjected to a
periodically oscillating magnetic field. We mainly focus on the
study of the two-dimensional system for various values of the
crystal-field coupling in the second-order transition regime. Our
results indicate that the present non-equilibrium phase transition
belongs to the universality class of the equilibrium Ising model
and allow us to construct a dynamic phase diagram, in analogy to
the equilibrium case, at least for the range of parameters
considered. Finally, we present some complementary results for the
three-dimensional model, where again the obtained estimates for
the critical exponents fall into the universality class of the
corresponding three-dimensional equilibrium Ising ferromagnet.
\end{abstract}

 \maketitle

\section{Introduction}
\label{sec:introduction}

Although our understanding of equilibrium critical phenomena has
developed to a point where well-established theories and results
are available for a wide variety of systems, far less is known for
the physical mechanisms underlying the non-equilibrium phase
transitions of many-body interacting systems. In this respect,
theoretical but also experimental studies deserve a particular
attention in order to provide further insight on the universality
and scaling principles of this type of phenomena. We know today
that when a ferromagnetic system, below its Curie temperature, is
exposed to a time-dependent oscillating magnetic field, it may
exhibit a fascinating dynamic magnetic behavior, which cannot be
directly obtained via its corresponding equilibrium
part~\cite{Tome}. In a typical ferromagnetic system being
subjected to an oscillating magnetic field, there occurs a
competition between the time scales of the applied field period
and the meta-stable lifetime, $\tau$, of the system. When the
period of the external field is selected to be smaller than
$\tau$, the time-dependent magnetization tends to oscillate around
a non-zero value, which corresponds to the dynamically ordered
phase. In this region, the time-dependent magnetization is not
capable to follow the external field instantaneously. However, for
larger values of the period of the external field, the system is
given enough time to indeed follow the external field. Hence, in
this case the time-dependent magnetization oscillates around its
zero value, indicating a dynamically disordered phase. When the
period of the external field becomes comparable to $\tau$, a
dynamic phase transition takes place between the dynamically
ordered and disordered phases.

Up to now, there have been several theoretical~\cite{Lo, Zimmer,
Acharyya1, Chakrabarti, Acharyya2, Acharyya3, Buendia1, Buendia2,
Jang1, Jang2, Shi, Punya, Riego, Keskin1, Keskin2, Robb1,Deviren,
Yuksel1, Yuksel2, Vatansever1}  and experimental~\cite{He, Robb,
Suen, Berger, Riego1} studies regarding dynamic phase transitions,
as well as the hysteresis properties of different types of
magnetic materials. These works indicate that, in addition to the
temperature, both the amplitude and the period of the
time-dependent magnetic field play a key role in dynamical
critical phenomena. On the other hand and to the best of our
knowledge, there exist only few studies focusing on the critical
exponents and universality aspects of spin models driven by a
time-dependent oscillating magnetic field~\cite{Sides1, Sides2,
Korniss, Buendia3, Park, Tauscher,Vatansever2}. In particular, by
means of Monte Carlo simulations and finite-size scaling analysis,
it has been suggested that the critical exponents of the
two-dimensional (2D) kinetic Ising model are compatible to those
of the corresponding 2D equilibrium Ising model~\cite{Korniss,
Sides1, Sides2}. In another relevant work~\cite{Buendia3},
Buend\'{i}a and Rikvold used soft Glauber dynamics to estimate the
critical exponents of the same system, providing strong evidence
that the characteristics of the phase transition are universal
with respect to the choice of the stochastic dynamics. The above
results have been corroborated by a numerical study of the
triangular-lattice Ising model~\cite{Vatansever2}, where the
obtained critical exponents were found to be consistent within
errors to those of the equilibrium Ising counterpart. Furthermore,
the universality features of the 3D kinetic Ising model have been
clarified by Park and Pleimling~\cite{Park}: the critical
exponents of 3D kinetic Ising model are in good agreement to those
of the corresponding equilibrium 3D case. Last but not least, the
role of surfaces at non-equilibrium phase transitions has been
elucidated in Ref.~\cite{Park2} where the non-equilibrium surface
exponents were found not to coincide with those of the equilibrium
critical surface and even more recently the fluctuations in a
square-lattice ferromagnetic model driven by a slowly oscillating
field with a constant bias have been studied in
Ref.~\cite{Buendia4}. This latter work, provided us with the
ubiquitous reminder that the equivalence of the dynamic phase
transition to an equilibrium phase transition is limited to the
critical region near the critical period and zero bias.

It is evident from the above discussion that most of the numerical
work performed to clarify the universality classes of dynamic
phase transitions has been devoted to the kinetic spin-$1/2$ Ising
type of models. Still, there is another suitable candidate model
where the above predictions may be tested: the so-called
Blume-Capel model~\cite{Capel, Blume}. The Blume-Capel model is
defined by a spin-1 Ising Hamiltonian with a single-ion uniaxial
crystal-field anisotropy (or simpler crystal-field coupling)
$\Delta$~\cite{Capel, Blume} [see also Eq.~(\ref{eq:1}) below].
The fact that this model has been very widely studied in
statistical and condensed-matter physics is explained not only by
its relative simplicity and the fundamental theoretical interest
arising from the richness of its phase diagram, but also by a
number of different physical realizations of variants of the
model~\cite{Lawrie,Selke}. From the theoretical point of view, in
order to have a better understanding of the equilibrium phase
transition characteristics, the model and its variants have been
intensively studied by making use of different methods, such as
renormalization-group calculations~\cite{Berker1, Branco,
Snowman}, Monte Carlo simulations~\cite{Jain, Falicov, Silva,
Malakis1, Malakis2, Malakis3, Fytas1, Kwak, Zierenberg}, and
mean-field theory approaches~\cite{Boccara, Hoston, Zahraouy}.

Despite intensive investigations devoted to the determination of
the time-dependent magnetic-field effects on the dynamic phase
transition nature of the spin-1 Blume-Capel model~\cite{Buendia1,
Keskin1, Keskin2, Deviren, ShiWei, Acharyya4}, critical exponents
and universality properties of the model have not been elucidated.
To fill this gap, we present in this paper the first study of
universality of the spin-1 square-lattice Blume-Capel model in the
neighborhood of a dynamic phase transition under the presence of a
time-dependent magnetic field. The aim of our study is threefold:
Firstly, we would like to check how the critical exponents of the
kinetic spin-1 Blume-Capel model, estimated at various values of
$\Delta$ in the second-order transition regime, compare to those
of the corresponding equilibrium Ising model. Secondly, we target
at constructing a dynamic phase diagram in the related plane for
the range of $\Delta$-values considered. In a nutshell, our
results indicate that the dynamic phase transition of the present
kinetic system belongs to the universality class of the
equilibrium Ising model. Some complementary results obtained for
the 3D version of this spin-1 kinetic model, presented at the end
of this paper, provide additional support in favor of this claim.
Furthermore, the obtained dynamic phase diagram is found to be
qualitatively similar to the equilibrium phase diagram constructed
in the crystal-field -- temperature plane~\cite{Silva, Malakis2,
Zierenberg}. Last but not least, the data given in this study
qualitatively support the previously published studies, where
general dynamic phase transition features of the same system have
been investigated via mean-field~\cite{Keskin1, Keskin2} and
effective-field~\cite{Deviren} theory treatments.

The outline of the remainder parts of the paper is as follows: In
Sec.~\ref{sec:model} we introduce the model and the details our
simulation protocol. In Sec.~\ref{sec:observables} we define the
relevant observables that will facilitate our finite-size scaling
analysis for the characterization of the universality principles
of this dynamic phase transition. The numerical results and
discussion for the 2D and 3D model are presented in
Secs.~\ref{sec:results_2D} and Sec.~\ref{sec:results_3D},
respectively. Finally, Sec.~\ref{sec:conclusions} contains a
summary of our conclusions.

\section{Model and Simulation Details}
\label{sec:model}

We consider the square-lattice Blume-Capel model under the
existence of a time-dependent oscillating magnetic field. The
Hamiltonian of the system reads as
\begin{equation}\label{eq:1}
 \mathcal{H} = -J\sum_{\langle xy \rangle}\sigma_{x}\sigma_{y}+\Delta \sum_{x}\sigma_{x}^2-h(t)\sum_{x}\sigma_{x} ,
\end{equation}
where the spin variable $\sigma_{x}$ takes on the values $-1$,
$0$, or $+1$, $\langle xy \rangle$ indicates summation over
nearest neighbors, and $J>0$ is the ferromagnetic exchange
interaction. $\Delta$ denotes the crystal-field coupling and
controls the density of vacancies ($\sigma_{x} = 0$). For $\Delta
\rightarrow -\infty$ vacancies are suppressed and the model
becomes equivalent to the Ising model. The term $h(t)$ corresponds
to a spatially uniform periodically oscillating magnetic field,
and, following the prescription of
Refs.~\cite{Korniss,Buendia3,Park}, we assume that all lattice
sites are exposed to a square-wave magnetic field with amplitude
$h_{0}$ and half period $t_{1/2}$.

The phase diagram of the equilibrium Blume-Capel model in the
crystal-field -- temperature plane consists of a boundary that
separates the ferromagnetic from the paramagnetic phase. The
ferromagnetic phase is characterized by an ordered alignment of
$\pm 1$ spins. The paramagnetic phase, on the other hand, can be
either a completely disordered arrangement at high temperature or
a $\pm1$-spin gas in a $0$-spin dominated environment for low
temperatures and high crystal fields. At high temperatures and low
crystal fields, the ferromagnetic-paramagnetic transition is a
continuous phase transition in the Ising universality class,
whereas at low temperatures and high crystal fields the transition
is of first-order character~\cite{Capel,Blume}. The model is thus
a classic and paradigmatic example of a system with a tricritical
point $[\Delta_{\rm t},T_{\rm t}]$~\cite{Lawrie}, where the two
segments of the phase boundary meet. A most recent reproduction of
the phase diagram of the model can be found in
Ref.~\cite{Zierenberg}, and an accurate estimation of the location
of the tricritical point has been given in Ref.~\cite{Kwak}:
$[\Delta_{\rm t},T_{\rm t}] = [1.9660(1), 0.6080(1)]$. However,
for the needs of the current work we restricted our analysis in
the second-order transition regime of the model $\Delta <
\Delta_{\rm t}$. In particular, we studied the system at the
following crystal-field values: $\Delta = 0$, $0.5$, $1$, $1.5$,
and $\Delta = 1.75$.

In numerical grounds, we performed Monte Carlo simulations on
square lattices with periodic boundary conditions using the
single-site update Metropolis
algorithm~\cite{Metropolis,Binder,Newman}. This approach, together
with the alternative option of stochastic Glauber
dynamics~\cite{Glauber:63}, consists the standard recipe in
kinetic Monte Carlo simulations, as was also noted in
Ref.~\cite{Buendia3}. In fact, very recently, the surface phase
diagram of the 3D kinetic Ising model in an oscillating magnetic
field has been studied within the framework of both Glauber and
Metropolis dynamics and it was been shown that the results remain
qualitatively unchanged when using different single-spin flip
dynamics~\cite{Tauscher}.

In our simulations, $N = L \times L$ defines the total number of
spins and $L$ the linear dimension of the lattice, taking values
within the range $L = 32 - 256$. For each pair of ($L$,
$\Delta$)-parameters we performed several independent long runs,
tailored for the value of $\Delta$ under study, using the
following protocol: the first $10^{3}$ periods of the external
field have been discarded during the thermalization process and
numerical data were collected and analyzed during the following
$10^{4}$ periods of the field. We note that the time unit in our
simulations is one Monte Carlo step per site (MCSS) and that error
bars have been estimated using the jackknife method~\cite{Newman}.
To set the temperature scale we fixed units by choosing $J=1$ and
$k_{\rm B}=1$, where $k_{\rm B}$ is the Boltzmann constant.
Appropriate choices of the magnetic-field strength, $h_{0} = 0.2$,
and the temperature, $T(\Delta) = 0.8 T_{\rm c}(\Delta)$, ensured
that the system is in the multi-droplet regime~\cite{Park}. Here,
$T_{\rm c}(\Delta)$ denotes the set of critical temperatures of
the equilibrium square-lattice Blume-Capel model, as estimated in
Ref.~\cite{Malakis2} and also given in Tab.~\ref{tab:1} below.
Finally, for the application of finite-size scaling on the
numerical data, we restricted ourselves to data with $L \geq
L_{\rm min}$. As usual, to determine an acceptable $L_{\rm min}$
we employed the standard $\chi^{2}$-test of goodness of
fit~\cite{Press}. Specifically, the $p$-value of our
$\chi^{2}$-test is the probability of finding an $\chi^{2}$ value
which is even larger than the one actually found from our data. We
considered a fit as being fair only if $10\% < p < 90\%$.

A similar prescription was also followed for the study of the 3D
version of the model and the details of this implementation will
be given later on, in the beginning of Sec.~\ref{sec:results_3D}.

\section{Observables}
\label{sec:observables}

In order to determine the universality aspects of the kinetic
Blume-Capel model, we shall consider the half-period dependencies
of various thermodynamic observables. The main quantity of
interest is the period-averaged magnetization
\begin{equation}\label{eq:2}
Q=\frac{1}{2t_{1/2}}\oint M(t)dt,
\end{equation}
where the integration is performed over one cycle of the
oscillating field. Given that for finite systems in the
dynamically ordered phase the probability density of $Q$ becomes
bimodal, one has to measure the average norm of $Q$ in order to
capture symmetry breaking, so that $\langle |Q| \rangle$ defines
the dynamic order parameter of the system. In the above
Eq.~(\ref{eq:2}), $M(t)$ is the time-dependent magnetization per
site
\begin{equation}\label{eq:3}
 M(t)=\frac{1}{N}\sum_{x = 1}^{N}\sigma_{x}(t).
\end{equation}

To characterize and quantify the transition using finite-size
scaling arguments we must also define quantities analogous to the
susceptibility in equilibrium systems. The scaled variance of the
dynamic order parameter
\begin{equation}\label{eq:4}
\chi_{L}^{Q} = N\left[\langle Q^2\rangle_{L} -\langle |Q|
\rangle^2_{L} \right],
\end{equation}
has been suggested as a proxy for the non-equilibrium
susceptibility, also theoretically justified via
fluctuation-dissipation relations~\cite{Robb1}. Similarly, one may
also measure the scaled variance of the period-averaged energy
\begin{equation} \label{eq:5}
\chi_{L}^{E} = N\left[\langle E^2\rangle_{L} -\langle E
\rangle^2_{L} \right],
\end{equation}
so that $\chi_{L}^{E}$ can be considered as the relevant heat
capacity of the dynamic system. Here $E$ denotes the
cycle-averaged energy corresponding to the cooperative part of the
Hamiltonian~(\ref{eq:1})
\begin{equation}
\label{eq:6} E=\frac{1}{2t_{1/2}N}\oint\left[-J\sum_{\langle xy
\rangle}\sigma_{x}\sigma_{y}+\Delta\sum_{x}\sigma_{x}^2\right]dt.
\end{equation}

A few comments are in order at this point with respect to the use
of the above Eqs.~(\ref{eq:5}) and (\ref{eq:6}), where we focus
only on the cooperative part of the energy in order to calculate
the time-averaged energy over a full cycle of the external field
and its corresponding variance. Conceptually, the role of the
time-averaged energy originating from an oscillating magnetic
field (namely, the time-dependent Zeeman term) can be better
understood with the help of the dynamic correlation function. In
spin systems driven by a time-dependent external field there may
be some dynamic correlations between the time-dependent magnetic
field and the time-dependent magnetization, which strongly depend
on the chosen temperature, including other parameters as well. In
order to explain this point in detail, let us define the dynamic
correlation function $G = \langle M(t)h(t) \rangle - \langle
M(t)\rangle \langle h(t)\rangle$, where $\langle \cdots \rangle$
denotes the time-average over a full cycle of the external
field~\cite{Acharyya95}. Since $\langle h(t) \rangle = 0$, we are
allowed to simplify as $G = \langle M(t)h(t) \rangle$. We know
that in the relatively strong ferromagnetic phase the spin-spin
interactions are dominant against the field energy. Therefore, the
spins do not tend to respond to the varying magnetic field for
fixed system parameters. In other words, the corresponding dynamic
correlation function is almost zero in this region. In the
regions, except from the strongly ferromagnetic and paramagnetic
phases, the relevant term may have a non-zero value, however the
energy term coming from this type of a behavior does not effect
the true dynamic phase transition point~\cite{Acharyya2}.

Finally, with the help of the dynamic order parameter $Q$ we may
define the corresponding fourth-order Binder
cumulant~\cite{Sides1,Sides2}
\begin{equation}
\label{eq:7}
U_{L}=1-\frac{\langle |Q|^4\rangle_L}{3\langle |Q|^2 \rangle_L^2},
\end{equation}
which provides us with an alternative estimation of the critical
point, giving at the same time a flavor of universality at its
intersection point~\cite{Binder81}.

\section{Results and discussion}
\label{sec:results_2D}

\begin{figure}[t]
\centering
\includegraphics[width=8 cm]{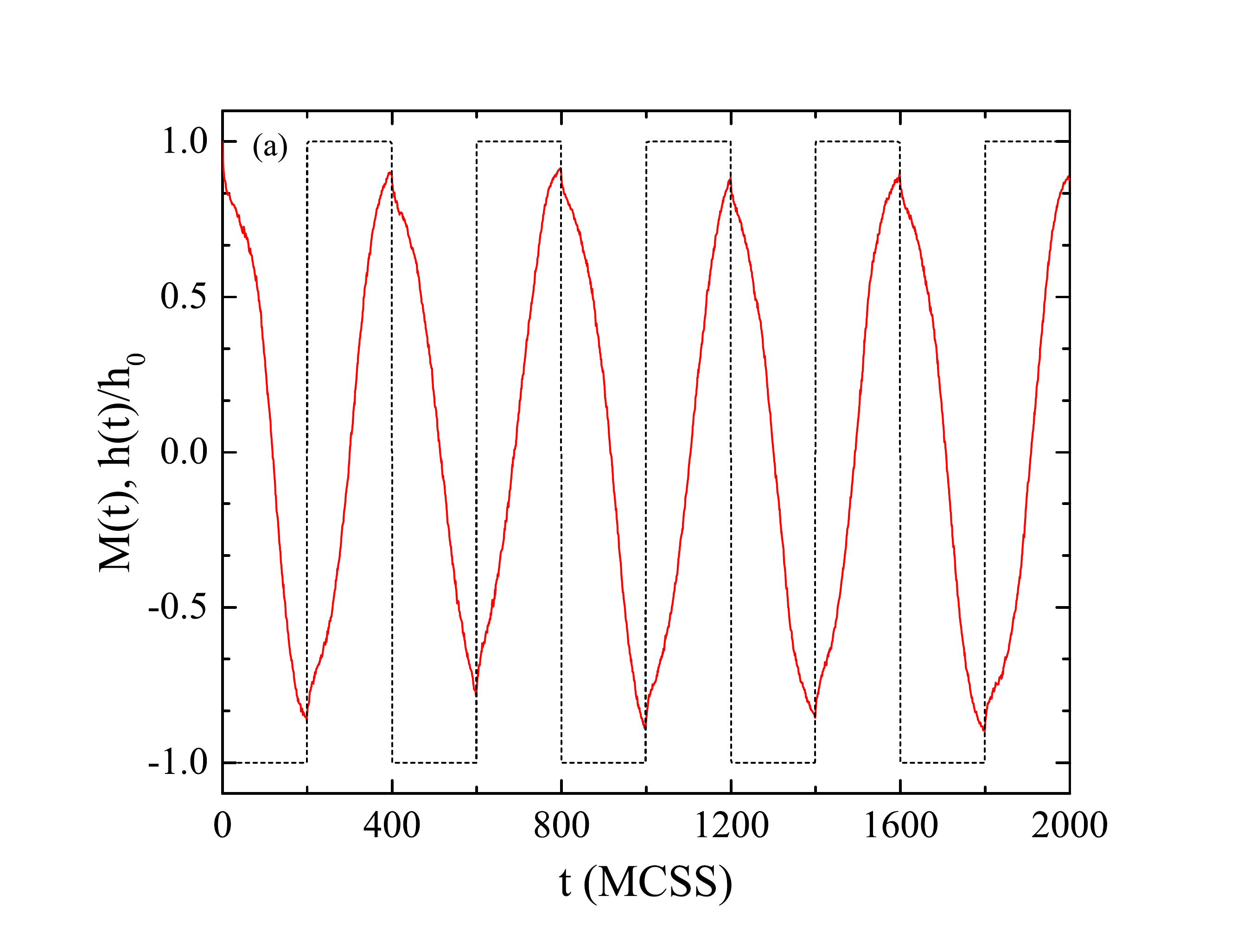}\\
\includegraphics[width=8 cm]{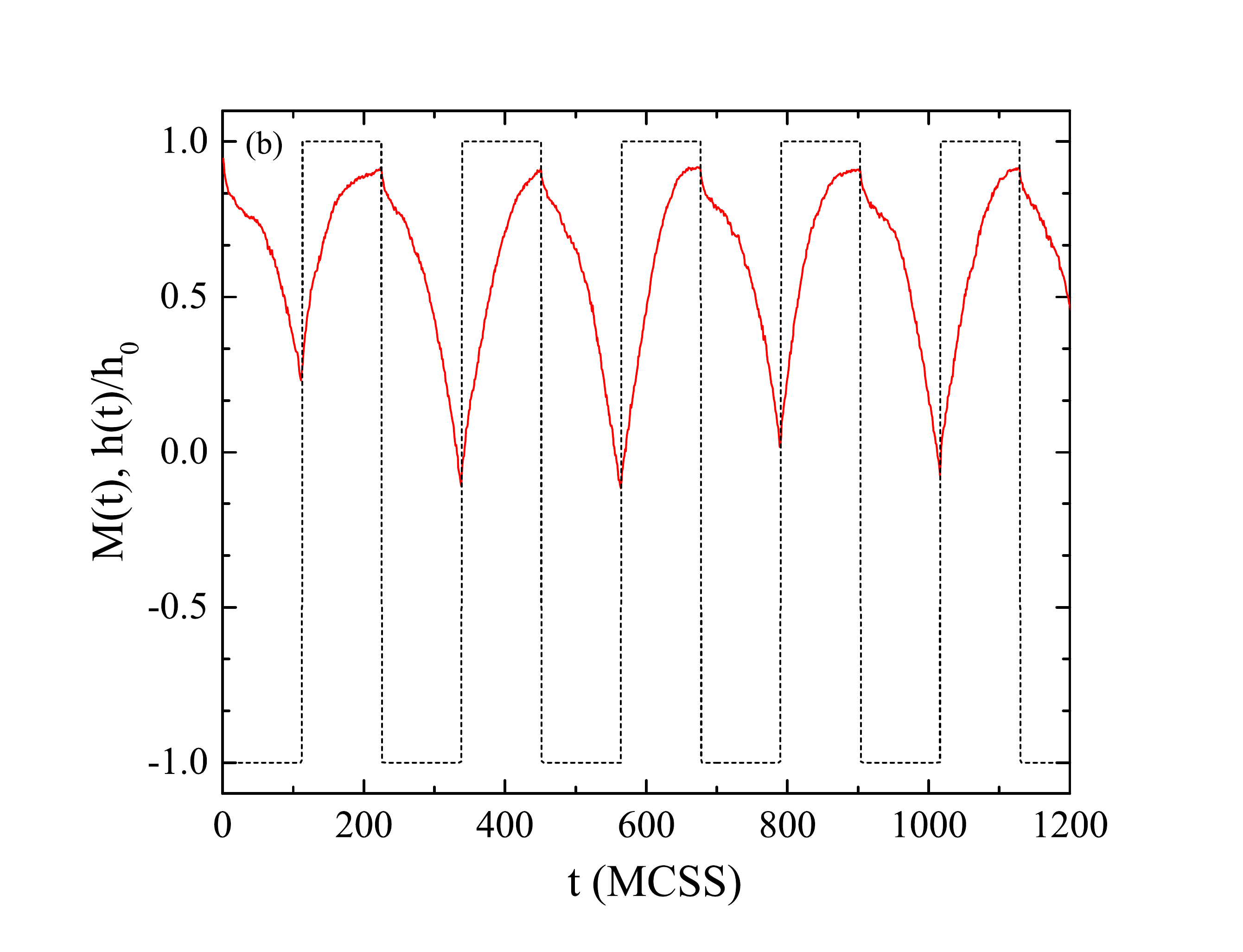}\\
\includegraphics[width=8 cm]{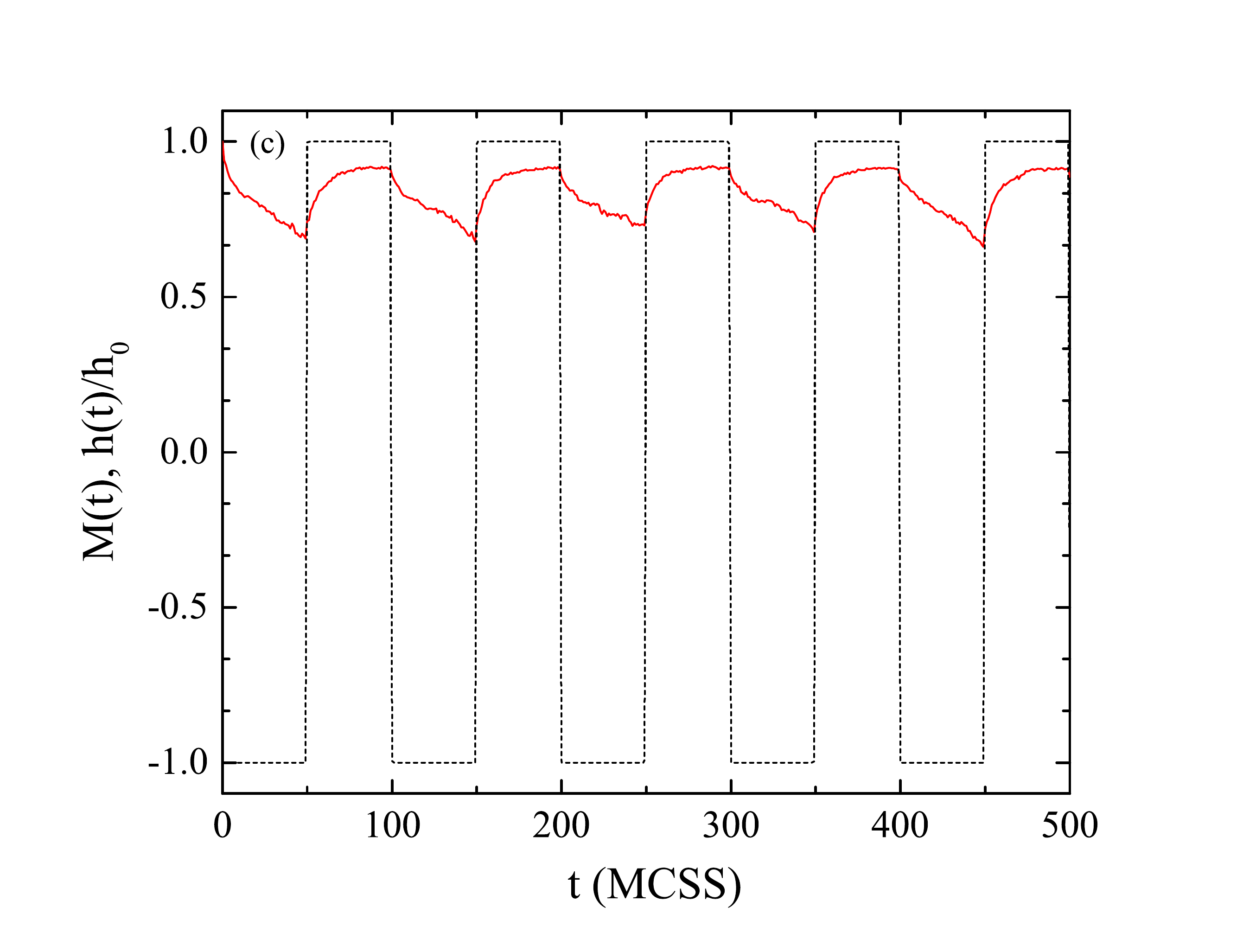}
\caption{\label{fig:1} (Color online) Time series of the
magnetization (red solid curves) of the $\Delta = 1$ kinetic
spin-1 Blume-Capel model under the presence of a square-wave
magnetic field (black dashed lines) for $L = 128$ and three values
of the half period of the external field: (a) $t_{1/2} = 200$
MCSS, corresponding to a dynamically disordered phase, (b)
$t_{1/2} = 113$ MCSS, close to the dynamic phase transition, and
(c) $t_{1/2} = 50$ MCSS, corresponding to a dynamically ordered
phase. Note that for the sake of clarity the ratio $h(t)/h_{0}$ is
displayed.}
\end{figure}

\begin{figure}[htbp]
\includegraphics*[width=12 cm]{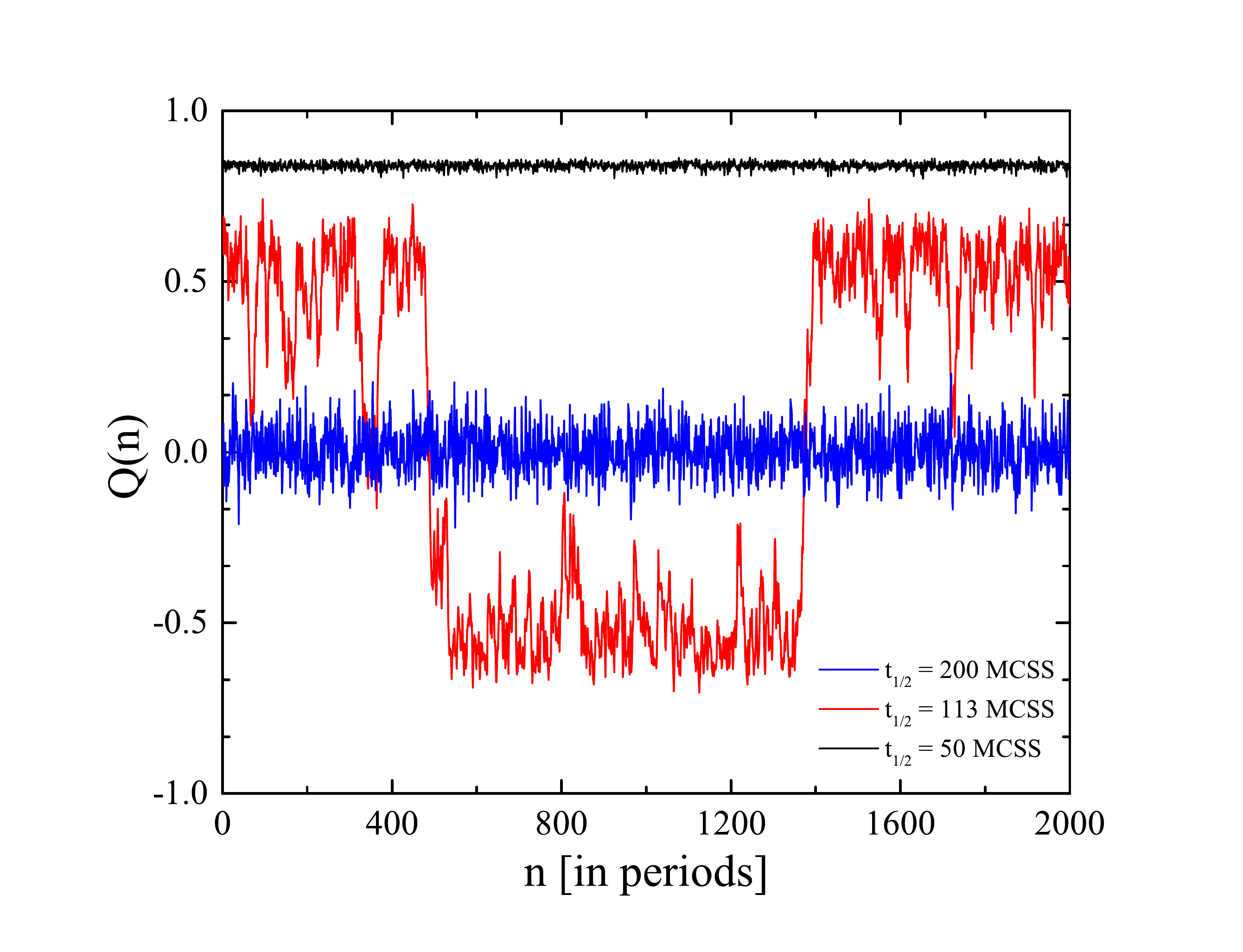}
\caption{\label{fig:2} (Color online) Period dependencies of the
dynamic order parameter of the $\Delta = 1$ kinetic spin-1
Blume-Capel model for $L = 128$. Results are shown for three
characteristic cases of the half period of the external field:
$t_{1/2} = 200$ MCSS (blue line), $t_{1/2} = 113$ MCSS (red line),
and $t_{1/2} = 50$ MCSS (black line). The strongly fluctuating
trace (red line) corresponds to the vicinity of the dynamic phase
transition, given that $t_{1/2} \approx t_{1/2}^{\rm c} = 112.3\pm
1.3$, as will be shown below.}
\end{figure}

It may be useful at this point to shortly describe the mechanism
underlying the dynamical ordering that takes place in kinetic
ferromagnets, as exemplified in Figs.~\ref{fig:1} and \ref{fig:2}
for the case of the $\Delta = 1$ Blume-Capel model and a system
size of $L = 128$. In particular, Fig.~\ref{fig:1} presents the
time evolution of the magnetization and Fig.~\ref{fig:2} the
period dependencies of the dynamic order parameter $Q$. Several
comments are in order at this point: For slowly varying fields,
Fig.~\ref{fig:1}(a), the magnetization follows the field,
switching every half period. In this region, as expected, $Q
\approx 0$, as also shown by the blue line in Fig.~\ref{fig:2}. On
the other hand, for rapidly varying fields, Fig.~\ref{fig:1}(c),
the magnetization does not have enough time to switch during a
single half period and remains nearly constant for many successive
field cycles, as also illustrated by the black line in
Fig.~\ref{fig:2}. In other words, whereas in the dynamically
disordered phase the ferromagnet is able to reverse its
magnetization before the field changes again, in the dynamically
ordered phase this is not possible and therefore the
time-dependent magnetization oscillates around a finite value. The
competition between the magnetic field and the meta-stable state
is captured by the half-period parameter $t_{1/2}$ (or by the
normalized parameter $\Theta = t_{1/2} /\tau$, with $\tau$ being
the meta-stable lifetime~\cite{Park}). Obviously, $t_{1/2}$ plays
the role of the temperature in the corresponding equilibrium
system. Now, the transition between the two regimes is
characterized by strong fluctuations in $Q$, see panel (b) in
Fig.~\ref{fig:1} and the evolution of the red line in
Fig.~\ref{fig:2}. This behavior is indicative of a dynamic phase
transition and occurs for values of the half period close to the
critical one $t_{1/2}^{\rm c}$ (otherwise stated when $t_{1/2}
\approx \tau$, so that $\Theta \approx 1$). Of course, since the
value $t_{1/2} = 113$ MCSS used for this illustration is slightly
above $t^{\rm c}_{1/2}$ for the case $\Delta = 1$, see
Tab.~\ref{tab:1}, the observed behavior includes as well some
non-vanishing finite-size effects.

\begin{figure}[t]
\centering
\includegraphics[width=8 cm]{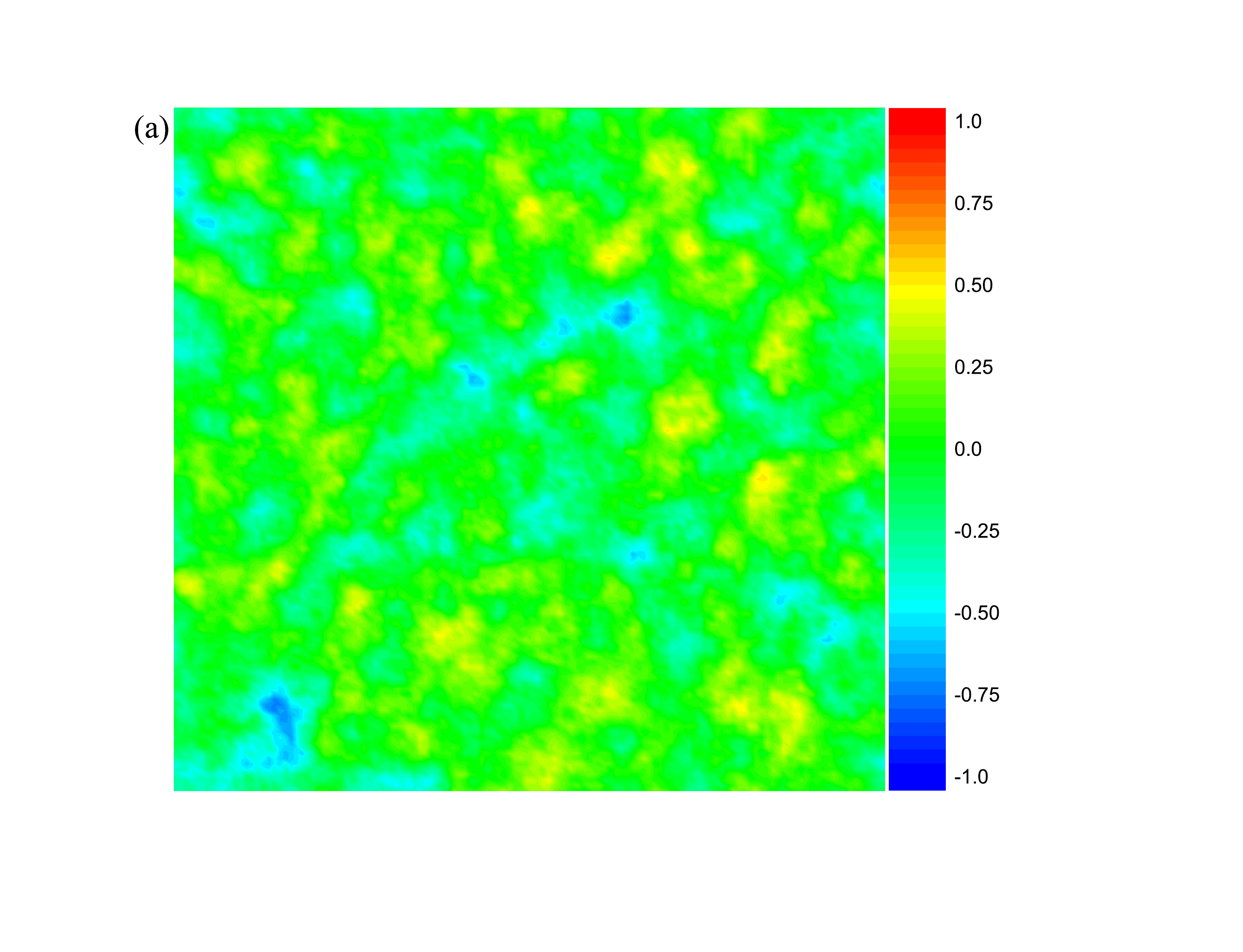}\\
\includegraphics[width=8 cm]{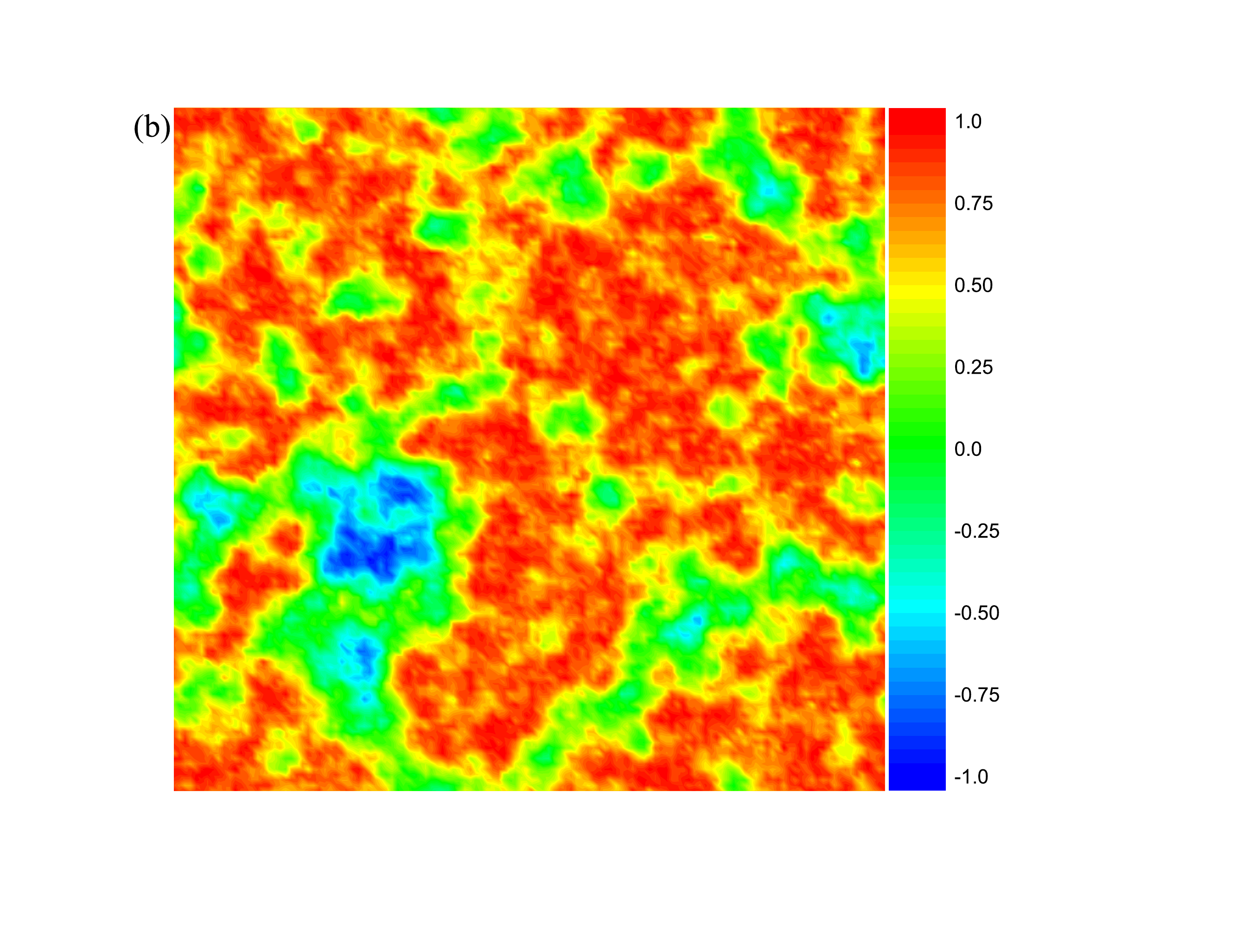}\\
\includegraphics[width=8 cm]{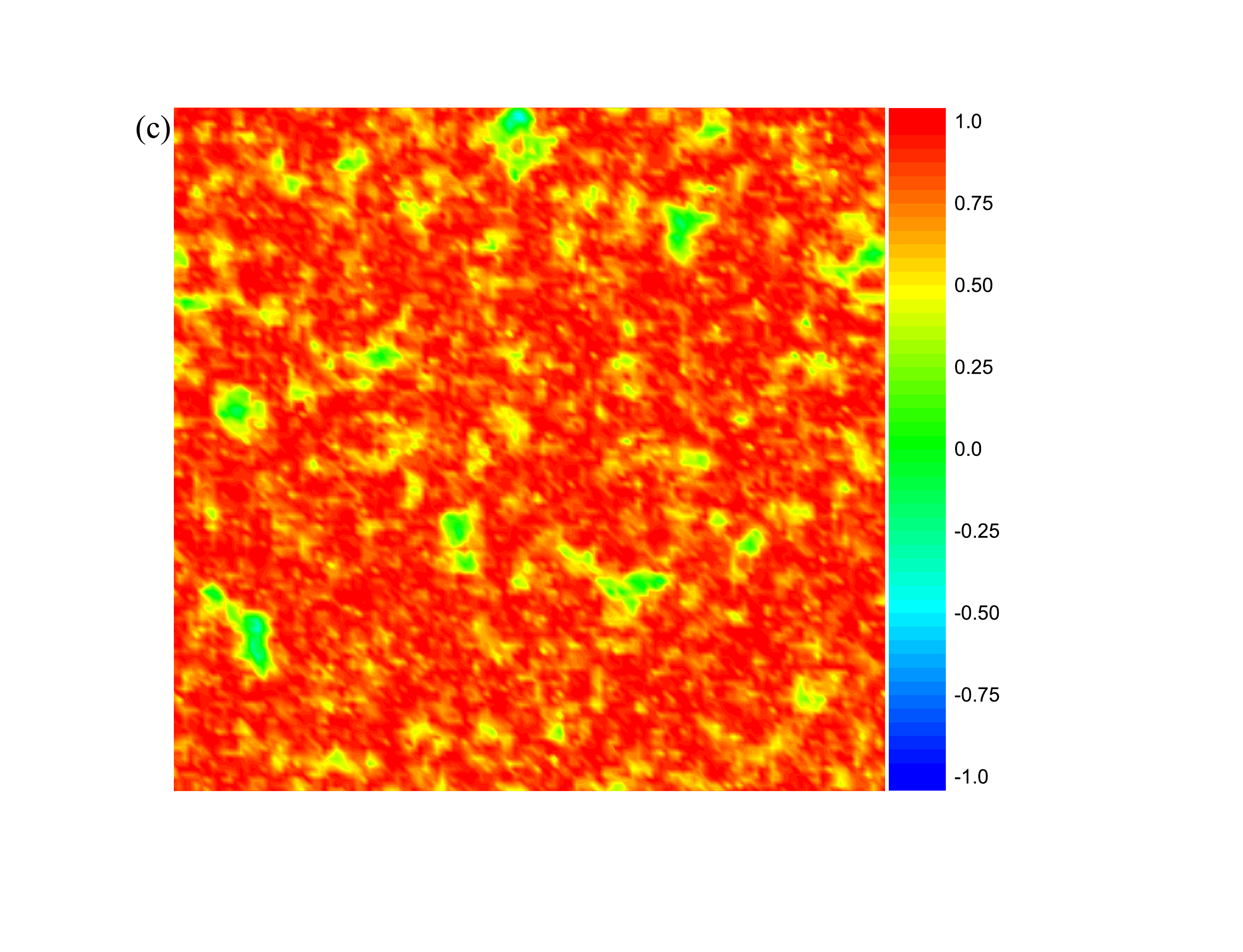}
\caption{\label{fig:3} (Color online) Configurations of the local
dynamic order parameter $\{Q_{x}\}$ of the $\Delta = 1.75$ kinetic
spin-1 Blume-Capel model for $L = 128$. The ``snapshots'' of
$\{Q_{x}\}$ for each regime are the set of local period-averaged
spins during some representative period. Three panels are shown:
(a) $t_{1/2} = 100$ MCSS $ > t_{1/2}^{\rm c}$ - dynamically
disordered phase, (b) $t_{1/2} = 43$ MCSS $\approx t_{1/2}^{\rm
c}$ - near the dynamic phase transition, and (c) $t_{1/2} = 20$
MCSS $ < t_{1/2}^{\rm c}$ - dynamically ordered phase.}
\end{figure}

To illustrate the spatial aspects of the transition scenario
described above in Figs.~\ref{fig:1} and \ref{fig:2}, we also show
configurations of the local order parameter $\{Q_{x}\}$ in
Fig.~\ref{fig:3} for the case of the $\Delta = 1.75$ Blume-Capel
model and a system of linear size $L = 128$. When the period of
the external field is selected to be bigger than the relaxation
time of the system, above $t_{1/2}^{\rm c}$, see panel (a), the
system follows the field in every half period, with some phase
lag, and $Q_{x}\approx 0$ at all sites $x$. In other words the
system lies in the dynamically disordered phase. On the other
hand, below $t_{1/2}^{\rm c}$, see panel (c), the majority of
spins spend most of their time in the $+1$ state, \emph{i.e.}, in
the meta-stable phase during the first half period, and in the
stable equilibrium phase during the second half period, except for
equilibrium fluctuations. Thus most of the $Q_{x}\approx +1$. The
system is now in the dynamically ordered phase. Near $t_{1/2}^{\rm
c}$ and the expected dynamic phase transition, there are large
clusters of both $+1$ and $-1$ states, within a sea of $0$-state
spins, as clearly illustrated in panel (b) of Fig.~\ref{fig:3}.

\begin{figure}[t]
\centering
\includegraphics[width=8 cm]{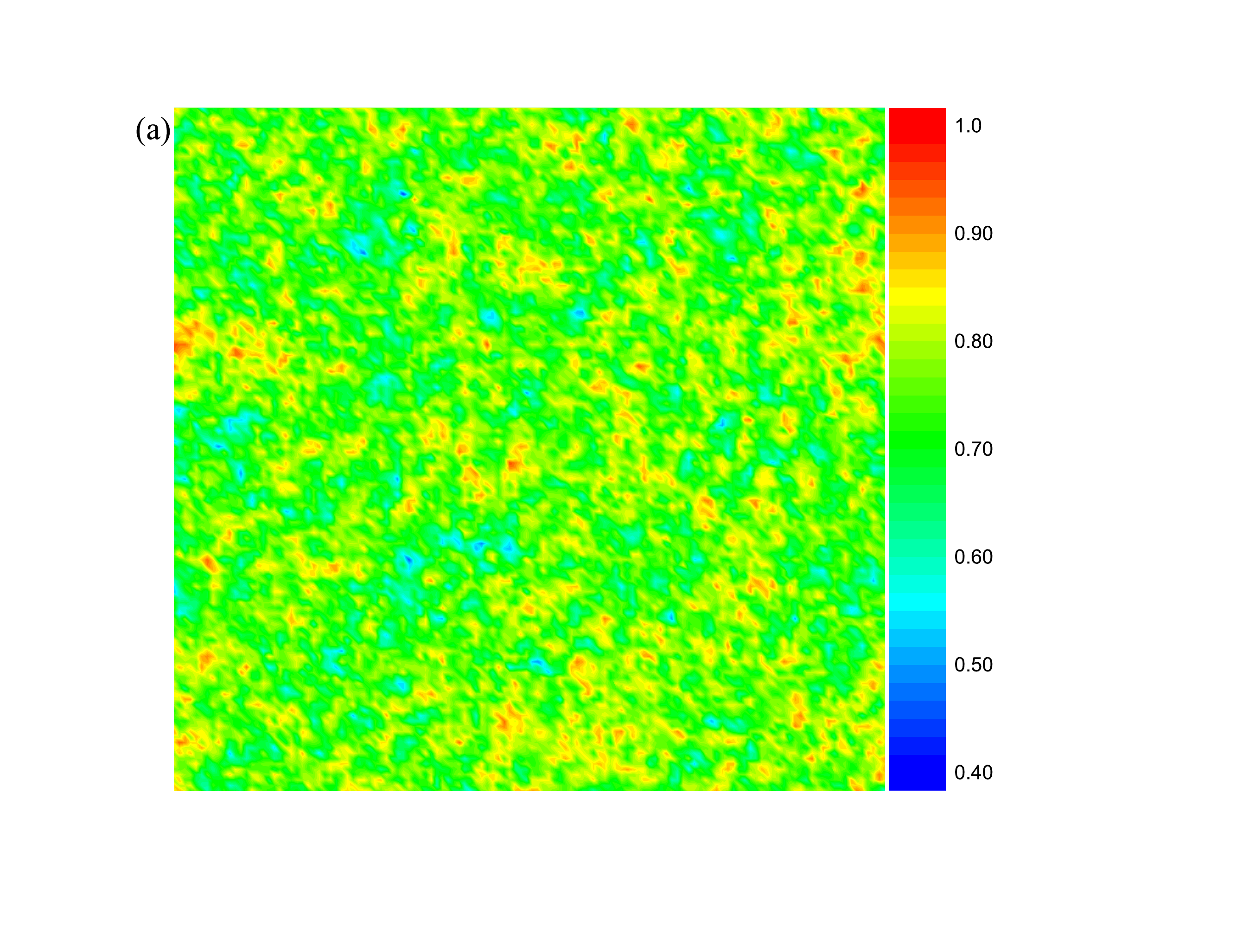}\\
\includegraphics[width=8 cm]{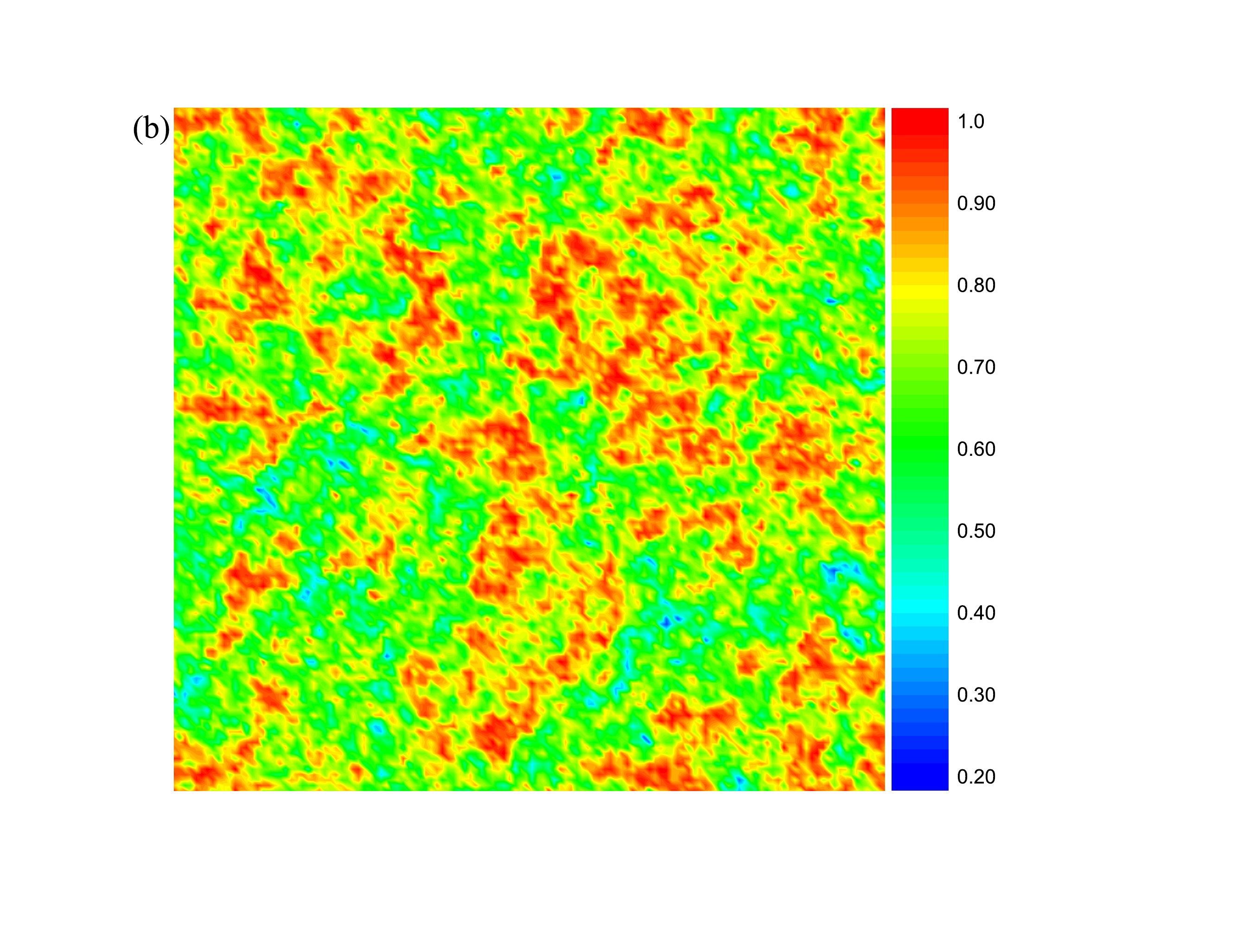}\\
\includegraphics[width=8 cm]{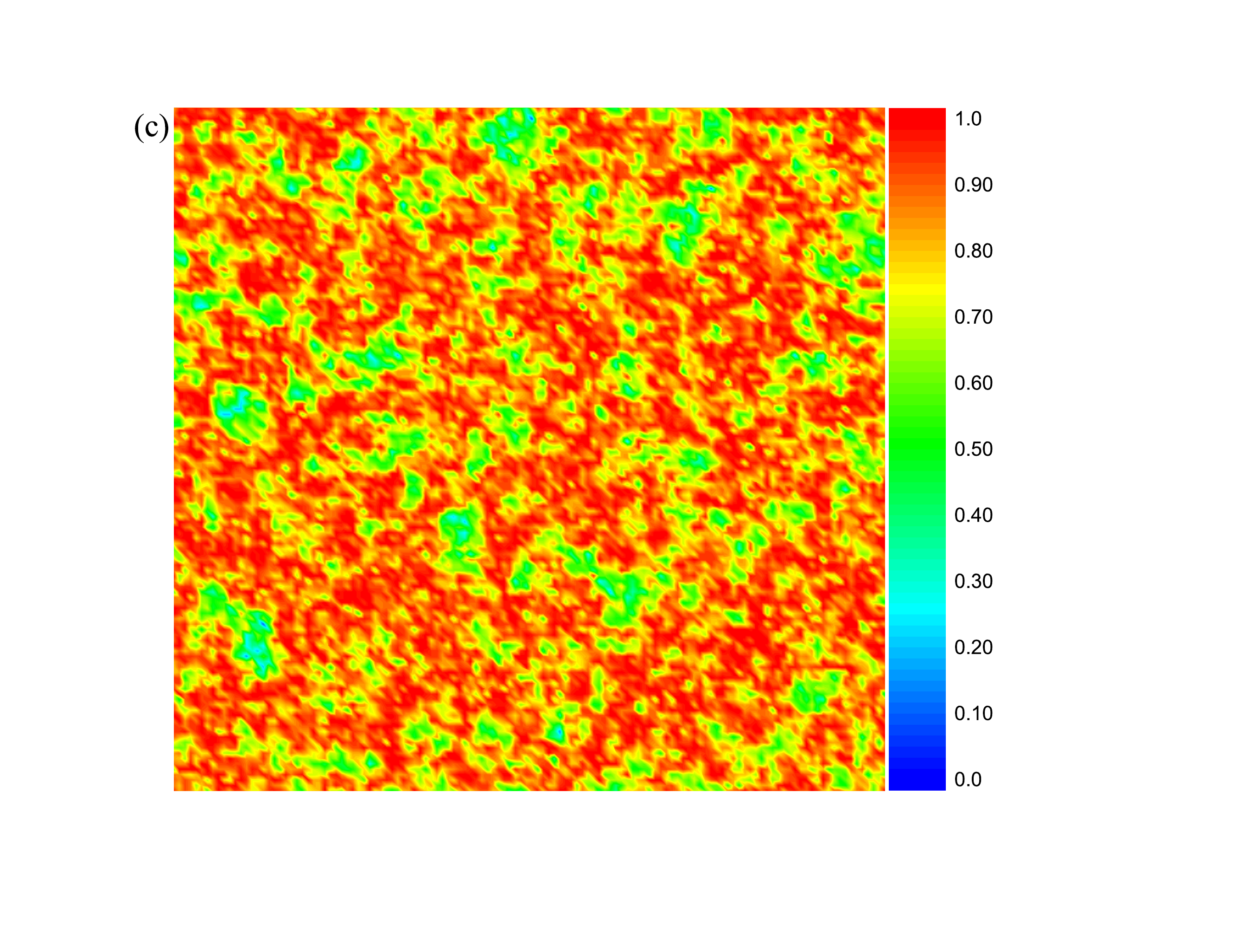}
\caption{\label{fig:4} (Color online) In full analogy to
Fig.~\ref{fig:3} we show snapshots of the period-averaged
quadrupole moment conjugate to the crystal-field coupling
$\Delta$. The simulation parameters are exactly the same as those
used in Fig.~\ref{fig:3} for all three panels (a) - (c).}
\end{figure}

However, the value of the local order parameter $\{Q_{x}\}$ does
not distinguish between random distributions of $\sigma_{x} = \pm
1$ and and regions of $\sigma_{x} = 0$. To bring out this
distinction, we present in Fig.~\ref{fig:4} similar snapshots of
the dynamic quadrupole moment over a full cycle of the external
field, $O = \frac{1}{2t_{1/2}}\oint q(t)dt$, where
$q(t)=\frac{1}{N}\sum_{x=1}^{N}\sigma_{x}^2$. The simulation
parameters are exactly the same to those used in Fig.~\ref{fig:3}
for all three panels (a) - (c). Of course, the dynamic quadrupole
moment is always one for the kinetic spin-1/2 Ising model, because
$\sigma_{x} = \pm 1$ in this case. In the spin-1 Blume-Capel model
the density of the ($\sigma_{x} = 0$) vacancies is controlled by
the the crystal-field coupling $\Delta$ and, thus, the value of
the dynamic quadrupole moment changes depending on $\Delta$. When
the  value of $\Delta$ increases, starting from its Ising limit
$(\Delta \rightarrow -\infty)$, the number of vacancies increases
as well in the system, so that the dynamic quadrupole moment tends
to decrease from its maximum value. In Fig.~\ref{fig:4}, except
from the regions with red color indicating the $+1$ state, the
regions enclosed by finite values exemplify the role played by the
the crystal-field coupling on the system.

To further explore the nature of the dynamic phase transition, we
performed a finite-size scaling analysis of the simulation data
obtained for various values of the crystal-field coupling
$\Delta$, as outlined above. Previous studies in the field
indicated that although scaling laws and finite-size scaling are
tools that have been designed for the study of equilibrium phase
transitions, they can be successfully applied as well to far from
equilibrium systems, like the current kinetic spin-1 Blume-Capel
model~\cite{Sides1,Sides2,Korniss,Buendia3,Park}.

\begin{figure}[htbp]
\includegraphics*[width=12 cm]{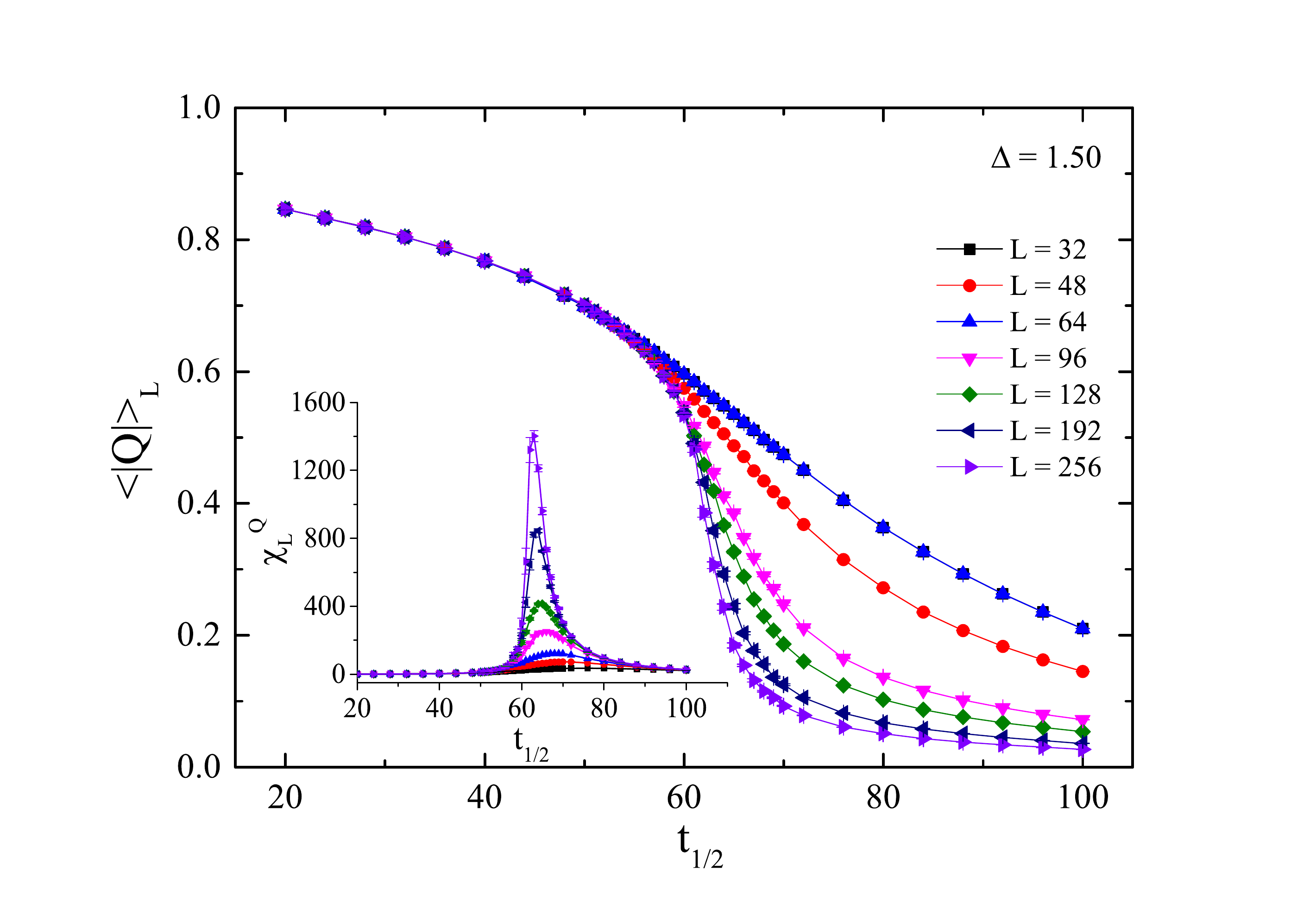}
\caption{\label{fig:5} (Color online) Half-period dependency of
the dynamic order parameter of the kinetic spin-1 $\Delta=1.5$
Blume-Capel model for a wide range of system sizes studied. The
inset represents the corresponding half-period dependency of the
corresponding dynamic susceptibility $\chi_{L}^{Q}$.}
\end{figure}

As an illustrative example for the case $\Delta = 1.5$, we present
in Fig.~\ref{fig:5} the finite-size behavior of the dynamic order
parameter (main panel) and the corresponding dynamic
susceptibility (inset) for a wide range of system sizes studied.
The main panel clearly shows that this dynamic order parameter
goes from a finite value to zero values at the half period
increases, showing a sharp change around the value of the half
period that can be mapped to the corresponding peak in the plot of
the dynamic susceptibility. The height and the location of the
maximum in $\chi_{L}^{Q}$ change with system size and we may
define as suitable pseudo-critical half periods these point
locations, denoted hereafter as $t_{1/2}^{\ast}$. The
corresponding maxima may be similarly defined as
$(\chi^{Q}_{L})^{\ast}$. Moreover, the absence of finite-size
effects below the critical point is a clear signature of a
divergent length scale. Of course, similar plots may be prepared
for all the other values of $\Delta$ studied, providing us with
suitable pseudo-critical points and susceptibility maxima that
will allow us to perform finite-size scaling.

\begin{table}
\caption{\label{tab:1}Summary of estimates for the critical half
period $t_{1/2}^{\rm c}$, the critical exponent $\nu$, and the
magnetic exponent ratio $\gamma/\nu$ of the spin-1 kinetic
Blume-Capel model for various values of the crystal-field coupling
$\Delta$, as illustrated. The second column lists the critical
temperatures of the equilibrium Blume-Capel model, as estimated in
Ref.~\cite{Malakis2}.}
\begin{ruledtabular}
\begin{tabular}{lcccc}
$\Delta$ & $T_{\rm c}$~\cite{Malakis2} & $t_{1/2}^{\rm c}$ & $\nu$ & $\gamma/\nu$ \\
\hline
0.00  & $1.693(3)$  & $206.4 \pm 1.2$ & 1.05(8) & 1.74(3)  \\
0.50  & $1.564(3)$  & $166.6 \pm 1.1$ & 1.01(7) &  1.75(1)  \\
1.00  & $1.398(2)$  & $112.3 \pm 1.3$ & 1.03(9) &  1.75(2)  \\
1.50  & $1.151(1)$  & $61.0 \pm 0.3$  & 0.98(6) &  1.76(1)  \\
1.75  & $0.958(1)$  & $43.1 \pm 0.2$  & 1.02(6) &  1.76(2)  \\
\end{tabular}
\end{ruledtabular}
\end{table}

\begin{figure}[htbp]
\includegraphics*[width=12 cm]{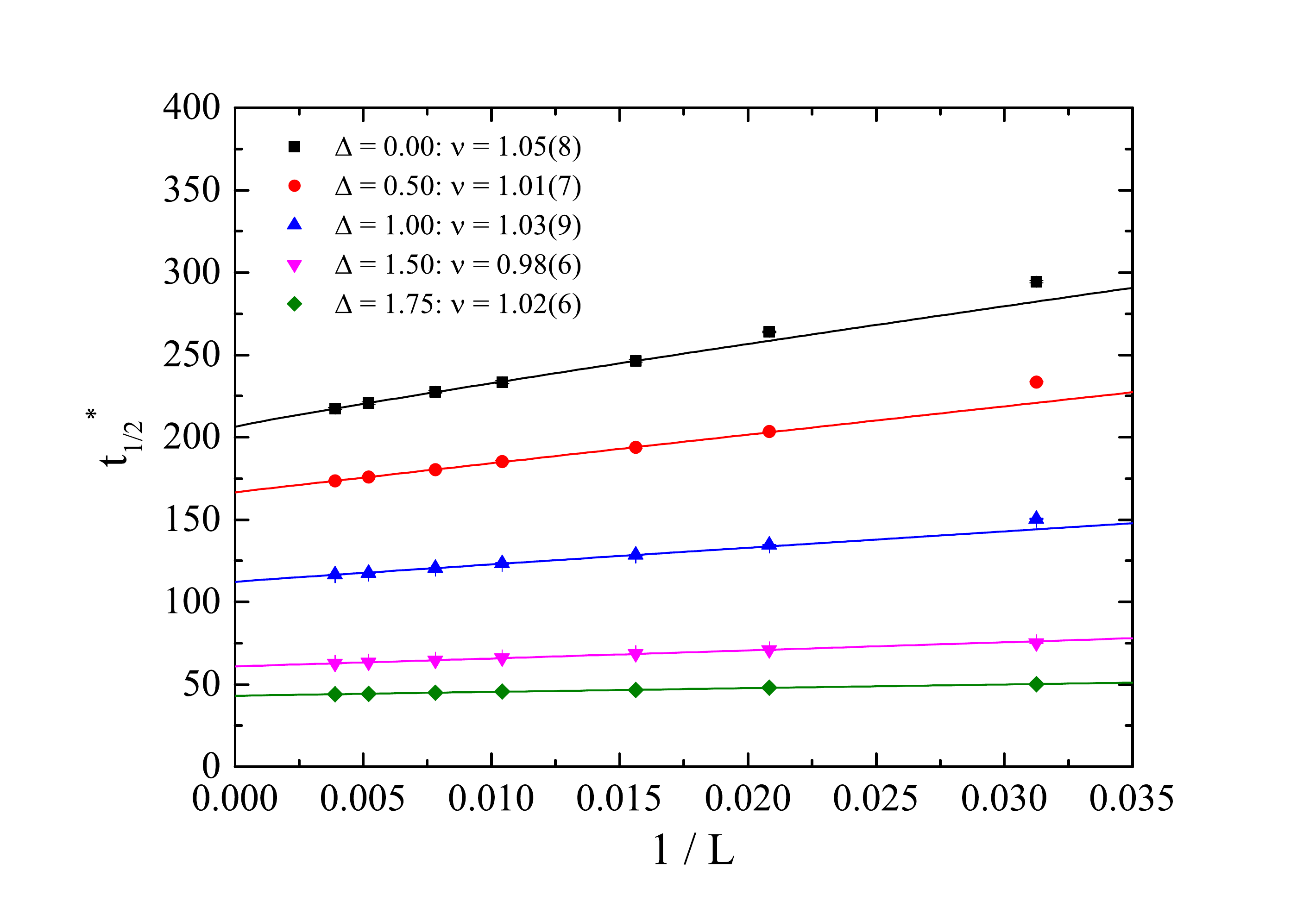}
\caption{\label{fig:6} (Color online) Estimation of the critical
half period $t_{1/2}^{\rm c}$ and the correlation-length's
exponent $\nu$ of the kinetic spin-1 Blume-Capel model for all
values of $\Delta$ considered. The solid lines are fits of
form~(\ref{eq:8}).}
\end{figure}

\begin{figure}[htbp]
\includegraphics*[width=12 cm]{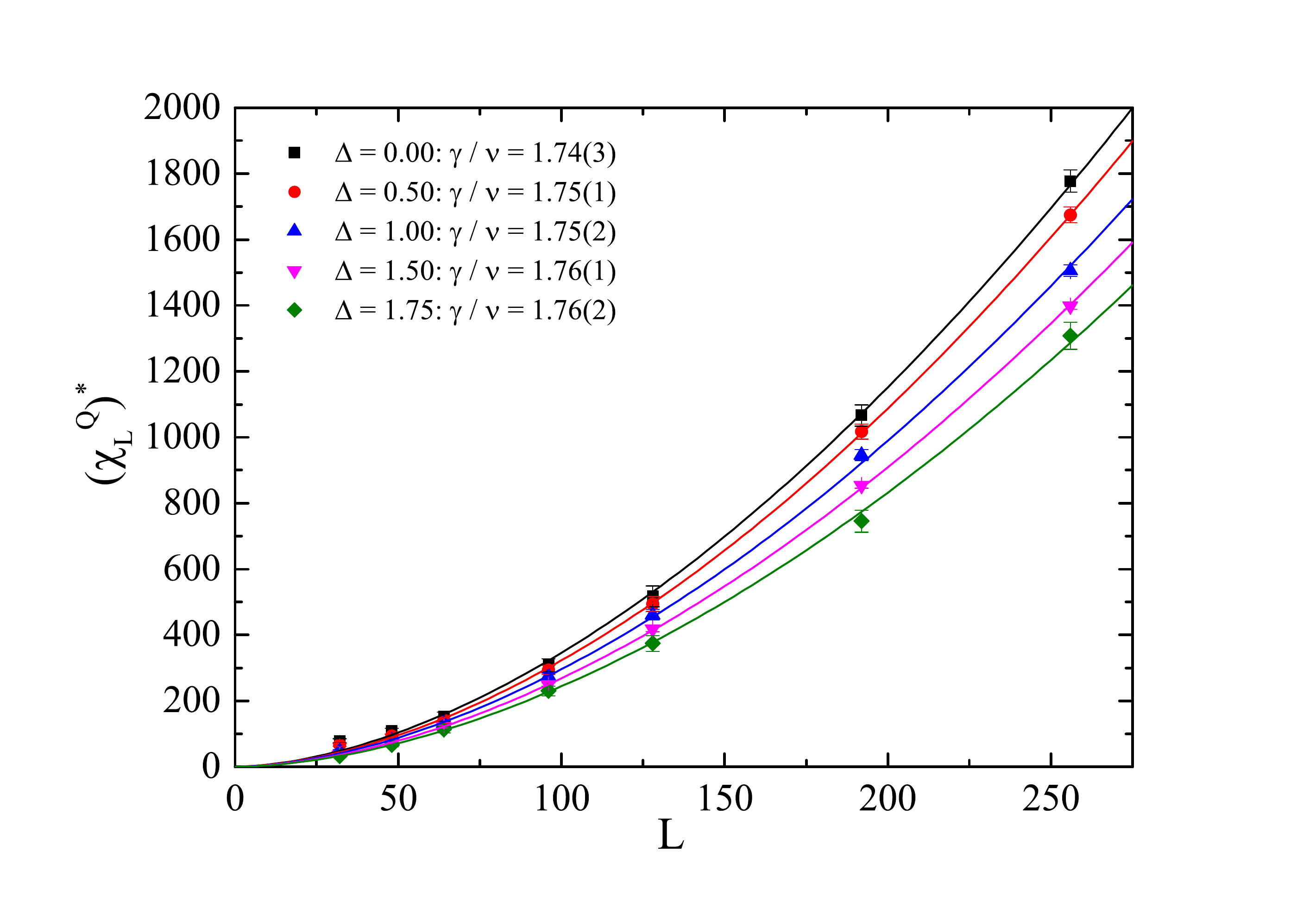}
\caption{\label{fig:7} (Color online) Finite-size scaling analysis
of the maxima $(\chi_{L}^Q)^{\ast}$ of the kinetic spin-1
Blume-Capel model for all values of $\Delta$ considered. The solid
lines are fits of the form~(\ref{eq:9}).}
\end{figure}

The shift behavior of the peak locations $t_{1/2}^{\ast}$ is
plotted in Fig.~\ref{fig:6} as a function of $1/L$ for all the
values of the crystal-field coupling considered. The solid lines
are fits of the usual shift form~\cite{Fisher,Privman,Binder92}
\begin{equation} \label{eq:8}
t_{1/2}^{\ast} = t_{1/2}^{\rm c} + bL^{-1/\nu},
\end{equation}
where $t_{1/2}^{\rm c}$ defines the critical half period of the
system and is a function of $\Delta$ and $\nu$ is the critical
exponent of the correlation length. The obtained values for the
critical half period are given at the third column of
Tab.~\ref{tab:1}. The relevant values for the critical exponent
$\nu$ are given in the panel of Fig.~\ref{fig:6} but are also
listed in the fourth column Tab.~\ref{tab:1}. These values suggest
that the critical exponent $\nu$ of the kinetic Blume-Capel model
is compatible, up to a very good accuracy, to the value $\nu = 1$
of the 2D equilibrium Ising model, thus providing a first strong
element of universality. Subsequently, in Fig.~\ref{fig:7} we
present the finite-size scaling analysis behavior of the peaks of
the dynamic susceptibility and the solid lines are fits of the
form~\cite{Ferrenberg}
\begin{equation}
\label{eq:9} (\chi^{Q}_{L})^{\ast} \sim L^{\gamma/\nu}.
\end{equation}
The results for the magnetic exponent ratio $\gamma/\nu$ are given
in the panel and also in the fifth column of Tab.~\ref{tab:1}.
Again, these values for all $\Delta$-cases studied in the present
work are in good agreement with the expected Ising value
$\gamma/\nu = 1.75$, reinforcing the scenario of universality for
the kinetic Blume-Capel model.

\begin{figure}[htbp]
\includegraphics*[width=12 cm]{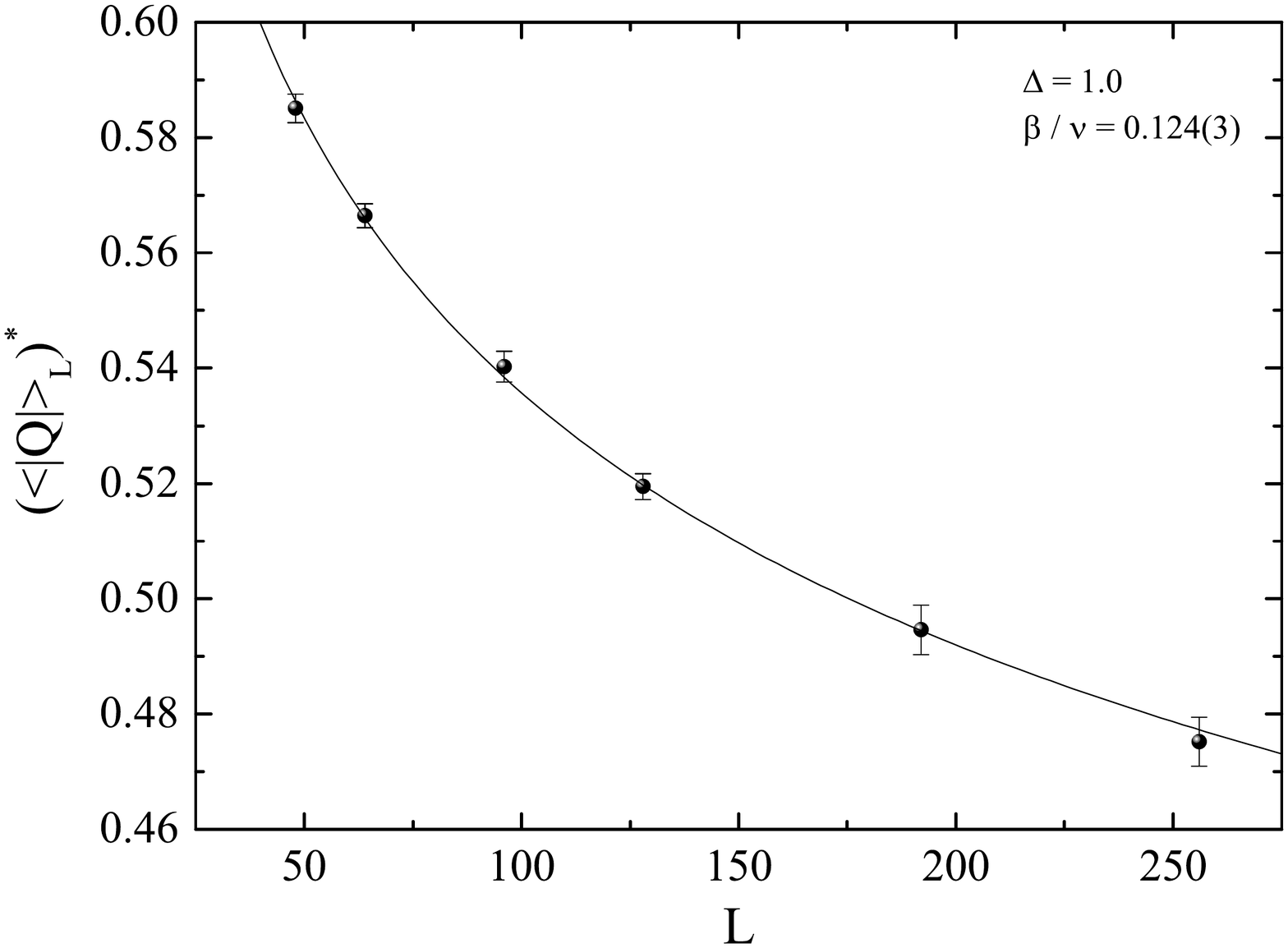}
\caption{\label{fig:8} Finite-size scaling analysis of the dynamic
order parameter estimated at the critical half period, $(\langle
|Q| \rangle_{L})^{\ast}$, of the 2D kinetic spin-1 Blume-Capel
model at $\Delta = 1$. The solid line is a power-law fit of the
form~(\ref{eq:10}).}
\end{figure}

In addition to $\gamma/\nu$, further evidence may be provided via
the alternative magnetic exponent ratio, namely $\beta/\nu$,
obtained from the scaling behavior of the dynamic order parameter
at the critical point via
\begin{equation}
\label{eq:10} (\langle |Q| \rangle_{L})^{\ast} \sim
L^{-\beta/\nu}.
\end{equation}
One characteristic example of this expected scaling behavior for
the kinetic spin-1 Blume-Capel model is shown in Fig.~\ref{fig:8}
for $\Delta = 1$. A power-law fit of the form~(\ref{eq:10}) gives
an estimate $0.124(3)$ for $\beta/\nu$, in good agreement with the
Ising value $1/8 = 0.125$. Let us note here that similar results
have been obtained in our fitting attempts for all the other
$\Delta$-values studied in this work.

\begin{figure}[htbp]
\includegraphics*[width=12 cm]{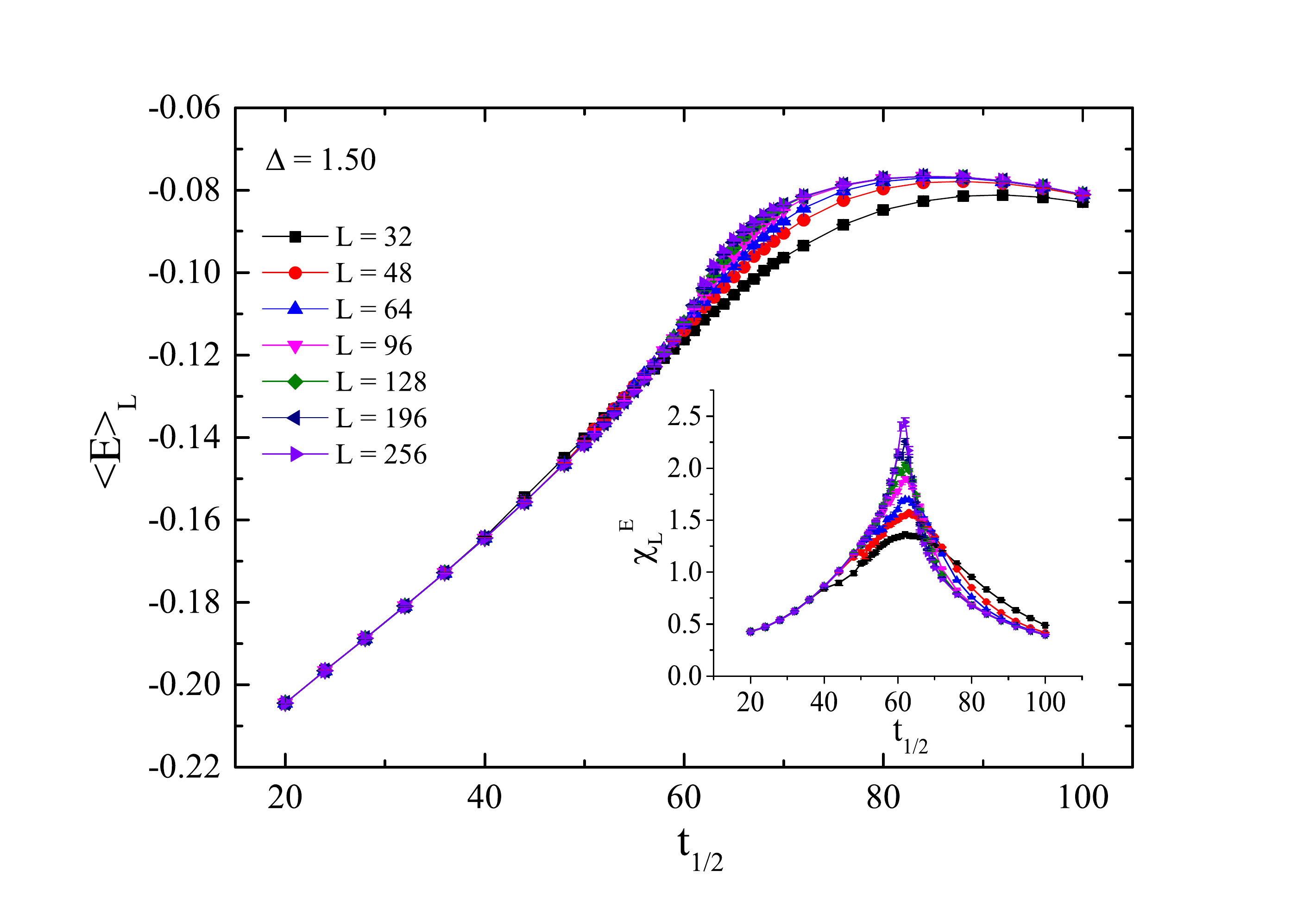}
\caption{\label{fig:9} (Color online) Half-period dependency of
the energy of the kinetic spin-1 $\Delta=1.5$ Blume-Capel model
for a wide range of system sizes studied. The inset illustrates
the half-period dependency of the corresponding heat capacity
$\chi_{L}^{E}$.}
\end{figure}

\begin{figure}[htbp]
\includegraphics*[width=12 cm]{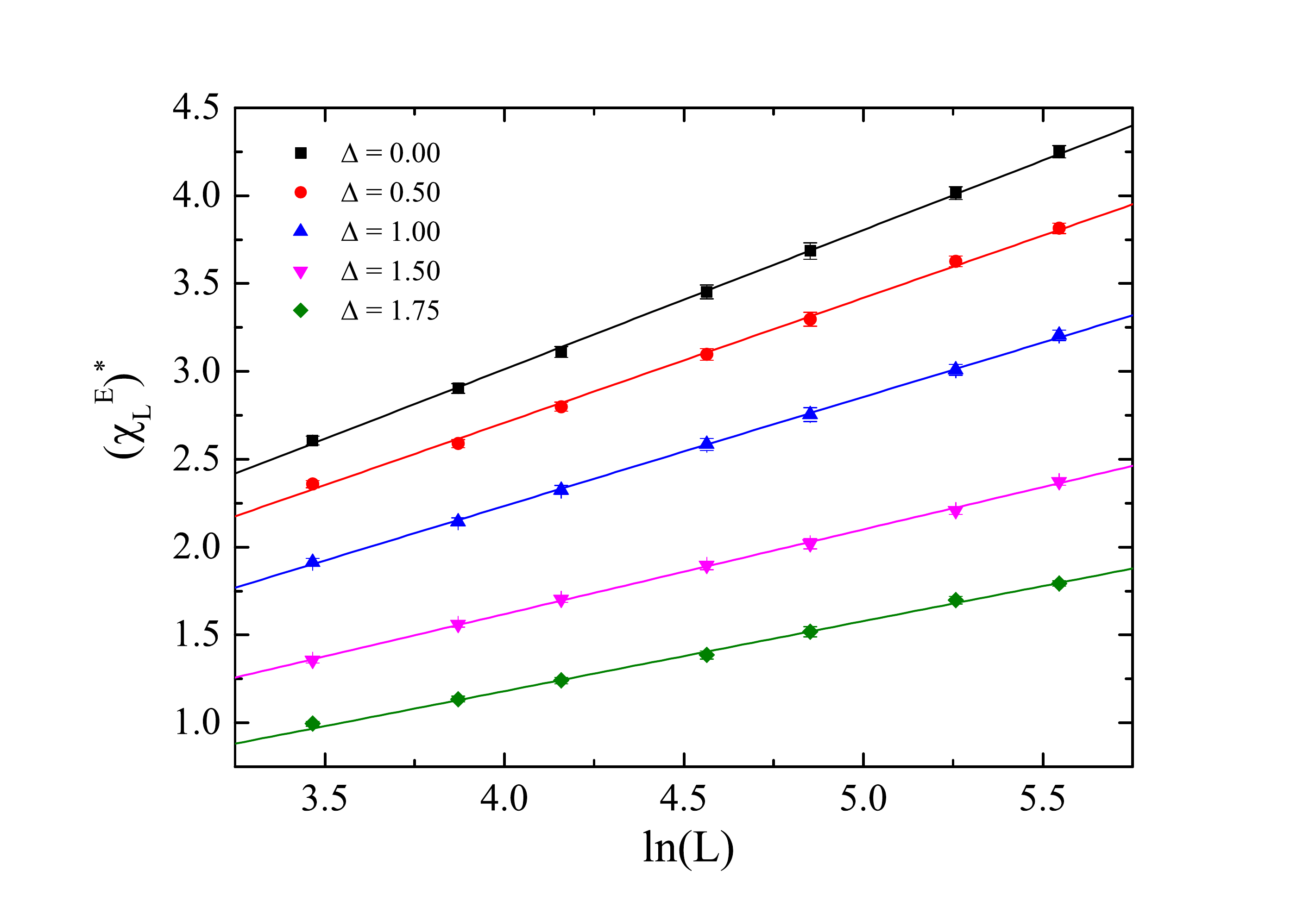}
\caption{\label{fig:10} (Color online) Illustration of the
logarithmic scaling behavior of the maxima of the heat capacity,
$(\chi_{L}^{E})^{\ast}$, for all values of $\Delta$ considered in
this work. The solid lines are fits of the form~(\ref{eq:11}).}
\end{figure}

As mentioned in Sec.~\ref{sec:observables}, we also measured the
energy and its corresponding scaled variance, the heat
capacity~(\ref{eq:5}). Both quantities are shown in the main panel
and the corresponding inset of Fig.~\ref{fig:9}, respectively.
Ideally, we would like to observe the logarithmic scaling behavior
of the maxima of the heat capacity $(\chi_{L}^{E})^{\ast}$.
Indeed, as it is shown in Fig.~\ref{fig:10}, the data for the
maxima of the heat capacity show a clear logarithmic divergence of
the form~\cite{Ferdinand}
\begin{equation}
\label{eq:11} (\chi^{E}_{L})^{\ast} \propto c_1+ c_2\ln{(L)},
\end{equation}
as expected for a 2D Ising ferromagnet.

\begin{figure}[htbp]
\includegraphics*[width=12 cm]{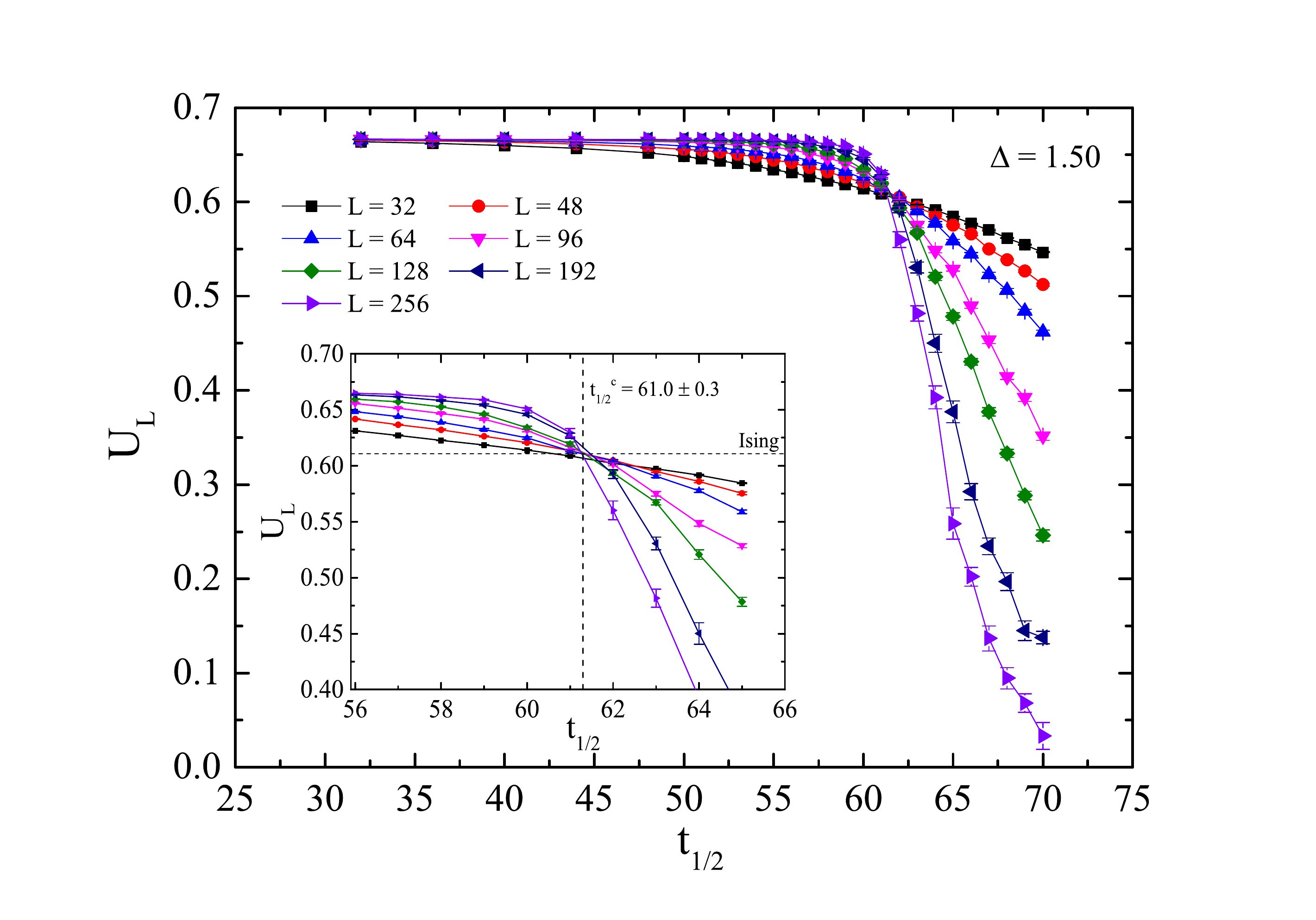}
\caption{\label{fig:11} (Color online) Half-period dependency of
the fourth-order Binder cumulant $U_{L}$ of the kinetic spin-1
$\Delta=1.5$ Blume-Capel model for a wide range of system sizes
studied. The inset displays an enhancement of the intersection
area.}
\end{figure}

A final verification of the equilibrium Ising universality class
comes form the study of the Binder cumulant, as defined above for
the case of the dynamic order parameter, see Eq.~(\ref{eq:7}). In
Fig.~\ref{fig:11} we plot the fourth-order cumulant $U_{L}$ for
the case $\Delta = 1.5$  for the various system sizes considered
in this work. The inset is a mere enlargement of the intersection
area. The vertical dashed line marks the critical half-period
value of the system $t_{1/2}^{\rm c}=61.0 \pm 0.3$, as estimated
in Fig.~\ref{fig:6}, and the horizontal dotted line marks the
universal value $U^{\ast} = 0.6106924(16)$ of the 2D equilibrium
Ising model~\cite{Salas}, which is consistent to the crossing
point of our numerical data. Certainly, the crossing point is
expected to depend on the lattice size $L$ (as it also shown in
the figure) and the term universal is valid for given lattice
shapes, boundary conditions, and isotropic interactions. For a
detailed discussion on this topic we refer the reader to
Refs.~\cite{Selke_2,Selke_3}. Still, the scope of the current
Fig.~\ref{fig:11} is to show qualitatively another instance of the
Ising universality. Similar plots and conclusions hold also for
the other values of $\Delta$ studied, but are omitted for brevity.
As a note, we remind the reader that an alternative way to
estimate the critical exponent $\nu$ comes from the scaling
behavior of the derivative of the Binder cumulant at the
corresponding crossing points, via $\left(\partial U_{L} /
\partial t_{1/2}\right) \propto L^{1/\nu}$, an approach that demands
quite accurate data at the area of the crossing points for a safe
estimation of derivatives~\cite{Binder}.

\begin{figure}[htbp]
\includegraphics*[width=12 cm]{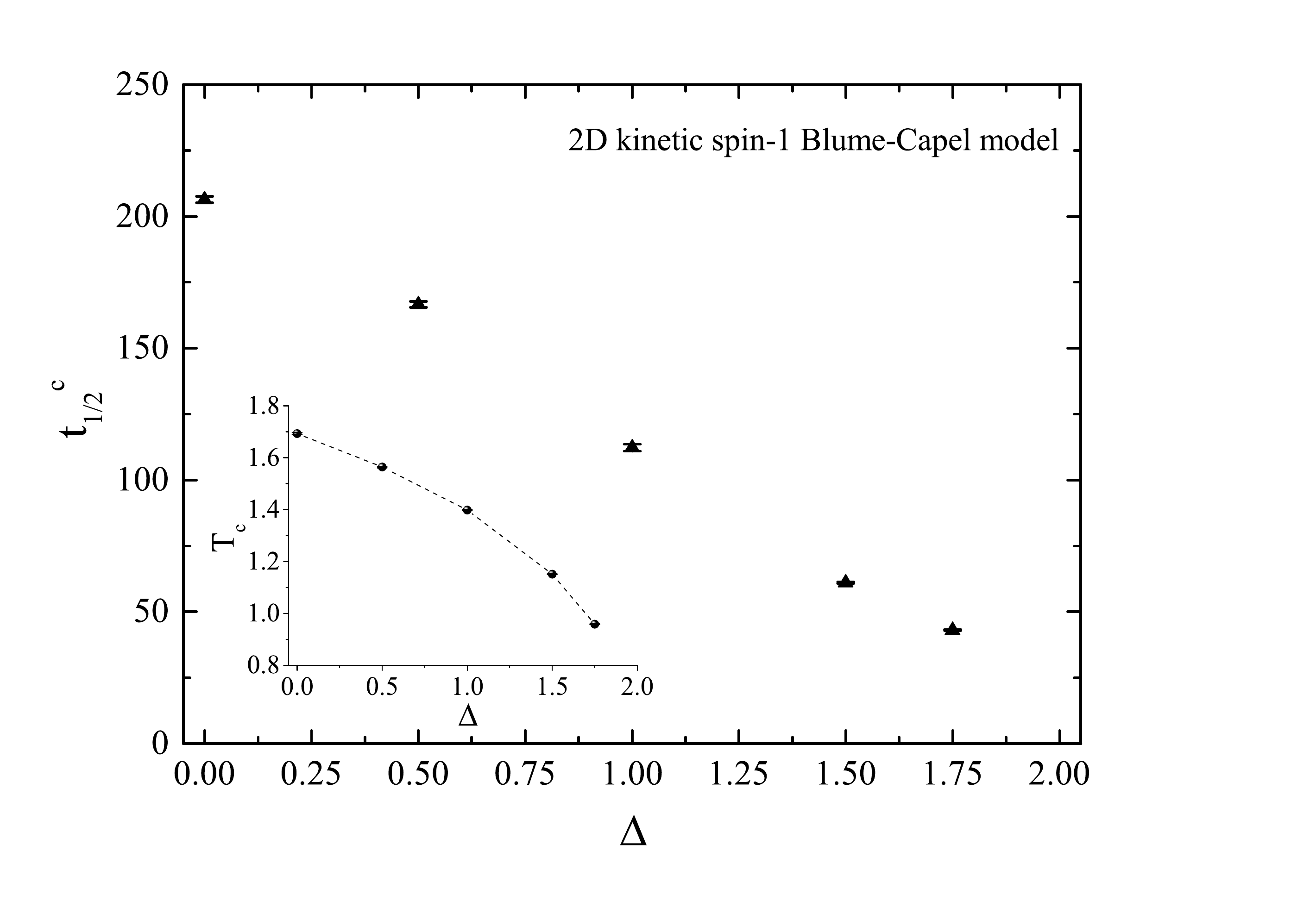}
\caption{\label{fig:12} Dynamic $\Delta-t_{1/2}^{\rm c}$ phase
boundary of the kinetic spin-1 square-lattice Blume-Capel model.
For the sake of completeness we provide in the inset a part of the
phase diagram of the equilibrium counterpart in the
$(\Delta-T_{\rm c})$ plane using the results of
Ref.~\cite{Malakis2} that were used in the present work. The
dotted line is a simple guide to the eye.}
\end{figure}

We complete our analysis on the 2D model, by presenting in
Fig.~\ref{fig:12} an illustrative formulation of a dynamic phase
diagram for the kinetic spin-1 Blume-Capel model at the
$(\Delta-t_{1/2}^{\rm c})$ plane, using the values for the
critical half period shown in Tab.~\ref{tab:1}. We also include a
complementary inset with the corresponding equilibrium counterpart
in the $(\Delta-T_{\rm c})$ plane using the results of
Ref.~\cite{Malakis2} for the regime of continuous transitions. As
it can be seen from the main panel of Fig.~\ref{fig:12}, the
values of $t_{1/2}^{\rm c}$ decrease almost linearly with
increasing $\Delta$. We are not currently sure if this is due to
the particular selection of the chosen temperatures, $0.8T_{\rm
c}(\Delta)$, or it may be a general result for any temperature
well below $T_{\rm c}$. Further simulations are needed to clarify
this point, that are however out of the scope of the current
study.

\section{On the dynamic phase transition of the 3D Blume-Capel model}
\label{sec:results_3D}

In this Section we present some complementary results on the
dynamic phase transition of the kinetic spin-1 3D Blume-Capel
model, as defined above in Eq.~(\ref{eq:1}), but with the spins
living on the simple cubic lattice. In this case $N = L\times L
\times L$, where $L\in\{8,16,24,32,48,64\}$. The analysis below is
presented for a single value of the crystal-field coupling in the
second-order transition regime of the phase diagram, namely for
$\Delta = 0.655$. For this value of $\Delta$, the critical
temperature of the model has been very accurately determined by
Hasenbusch to be $T_{\rm c} = 2.579169$~\cite{Hasenbusch}. Our
Monte Carlo simulations followed the protocol defined above in
Sec.~\ref{sec:model} for the case of the square-lattice model,
using now $h_0 = 0.6$ and $T = 0.8 T_{\rm c}$ as appropriate
choices for the magnetic-field strength and the temperature,
respectively.

We summarize our results in Figs.~\ref{fig:13} - \ref{fig:15}
below. In particular, in the main panel of Fig.~\ref{fig:13} we
present the finite-size scaling behavior of the maxima
$(\chi_{L}^Q)^{\ast}$. The solid line in is a fit of the
form~(\ref{eq:9}) providing us with the estimate $\gamma/\nu =
1.97(2)$ which is in very good agreement to the Ising value
$1.96370(2)$ of the equilibrium 3D Ising ferromagnet~\cite{Kos}.
In the corresponding inset we illustrate the shift behavior of the
pseudo-critical half periods $t_{1/2}^{\ast}$ as a function of
$L^{-1/\nu}$, where $\nu$ has been fixed to the Ising value
$0.629971$~\cite{Kos}. The numerical data are well described by a
linear extrapolation to the $L\rightarrow \infty$ limit, see also
Eq.~(\ref{eq:8}), indicating that the critical exponent $\nu$ of
the kinetic version of the 3D Blume-Capel model shares the value
of its equilibrium counterpart. The critical half period is also
estimated to be $t_{1/2}^{\rm c} = 72.1\pm 0.6$ for the particular
case of $\Delta$ studied here, as also indicated in the inset.

Figure~\ref{fig:14} illustrates the scaling behavior of the
dynamic order parameter of the 3D kinetic spin-1 Blume-Capel model
at the above estimated critical half period, $(\langle |Q|
\rangle_{L})^{\ast}$. Similarly to the 2D model, the solid line is
a power-law fit of the form~(\ref{eq:10}), providing the estimate
$\beta / \nu = 0.516(37)$ for the magnetic exponent ratio that
should be compared to the value $0.518149(6)$ of the 3D Ising
model~\cite{Kos}. Again, the agreement is beyond any (numerical)
doubt.

Finally we discuss the scaling behavior of the heat
capacity~(\ref{eq:5}). In Fig.~\ref{fig:15} we present the size
evolution of the maxima of the heat capacity
$(\chi_{L}^{E})^{\ast}$ of the 3D kinetic spin-1 Blume-Capel
model. The solid line is a power-law fit of the
form~\cite{Privman}
\begin{equation}
\label{eq:12} (\chi^{E}_{L})^{\ast} \sim L^{\alpha/\nu},
\end{equation}
and the obtained estimate for the critical exponent ratio
$\alpha/\nu = 0.17(2)$ nicely compares to the equilibrium value
$\alpha/\nu = 0.17475(2)$ of the 3D Ising ferromagnet~\cite{Kos},
thus supporting the conclusion of an earlier work by Park and
Pleimling on the 3D kinetic Ising model~\cite{Park}.

\section{Summary}
\label{sec:conclusions}

In the present work we investigated the dynamical response of the
2D Blume-Capel model exposed to a square-wave oscillating external
field. Using Monte Carlo simulations and finite-size scaling
techniques we studied the system at various values of the
crystal-field coupling within the second-order transition regime.
Our results for the critical exponent $\nu$, the magnetic exponent
ratios $\gamma/\nu$ and $\beta/\nu$, the universal Binder
cumulant, as well as the observed logarithmic divergence of the
heat capacity, indicate that the present non-equilibrium phase
transition belongs to the universality class of the equilibrium
Ising model. Furthermore, with the numerical data at hand, we have
been able to construct a 2D dynamic phase diagram for the range of
parameters considered, in analogy to the equilibrium case.
Additional evidence in favor of this universality scenario between
the dynamic phase transition and its equilibrium counterpart was
given via a supplemental study of the 3D Blume-Capel model.

To conclude, the results presented in the current paper, together
with existing ones for the 2D and 3D kinetic Ising
models~\cite{Korniss, Sides1,
Sides2,Buendia3,Park,Tauscher,Vatansever2} establish a clear
universality between the equilibrium and dynamic phase transitions
of Ising type of spin models. They also provide additional credit
to the symmetry arguments put forward by Grinstein \emph{et
al.}~\cite{Grinstein} a few decades ago, underlying the role of
symmetries in non-equilibrium critical phenomena.

\begin{acknowledgments}
The authors would like to thank Per Arne Rikvold and Walter Selke
for many useful comments on the manuscript. The numerical
calculations reported in this paper were performed at
T\"{U}B\.{I}TAK ULAKBIM (Turkish agency), High Performance and
Grid Computing Center (TRUBA Resources).
\end{acknowledgments}

\begin{figure}[htbp]
\includegraphics*[width=12 cm]{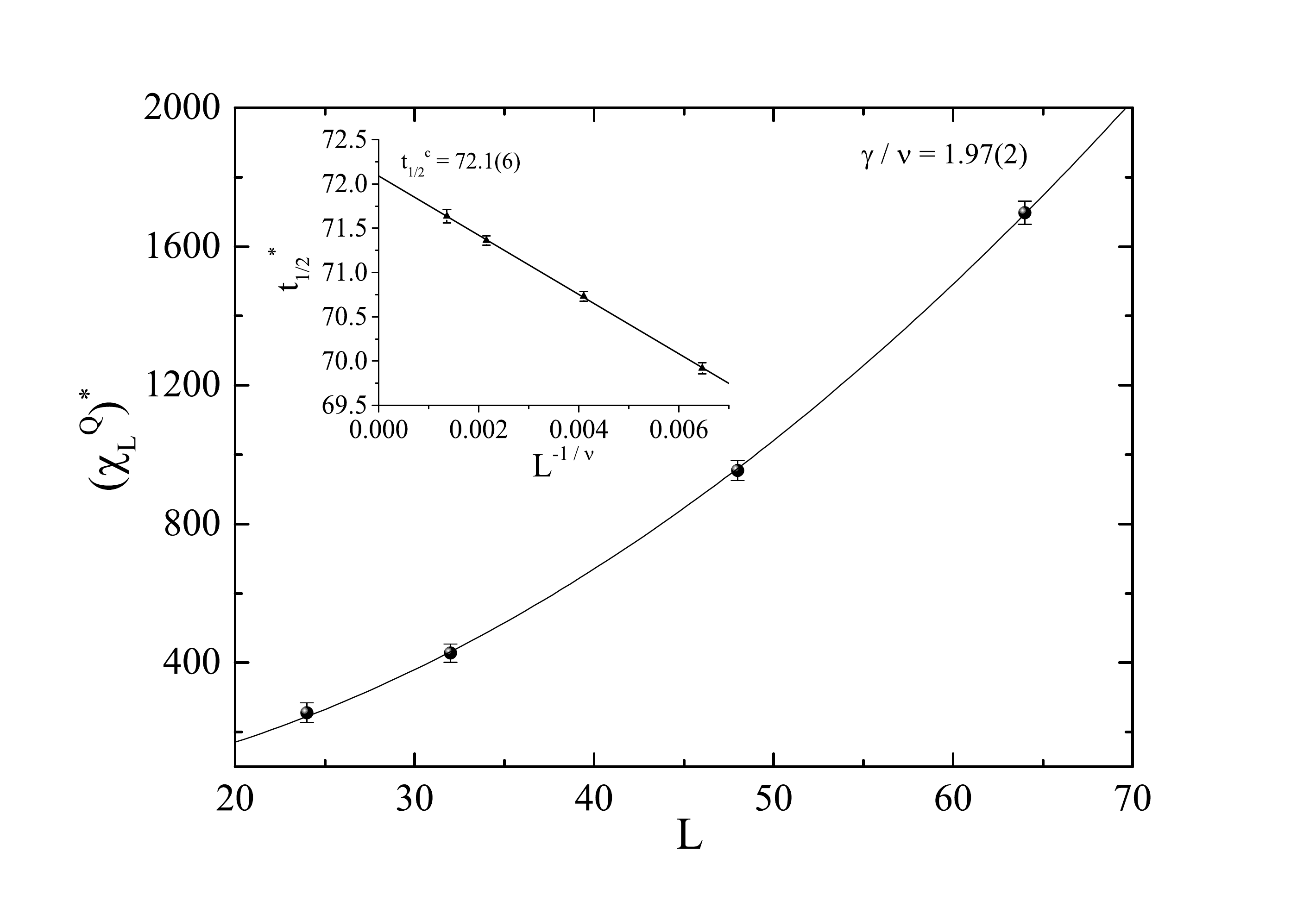}
\caption{\label{fig:13} Finite-size scaling analysis of the maxima
$(\chi_{L}^Q)^{\ast}$ (main panel) and shift behavior of the
corresponding pseudo-critical half periods $t_{1/2}^{\ast}$
(inset) of the 3D kinetic spin-1 Blume-Capel model.}
\end{figure}

\begin{figure}[htbp]
\includegraphics*[width=12 cm]{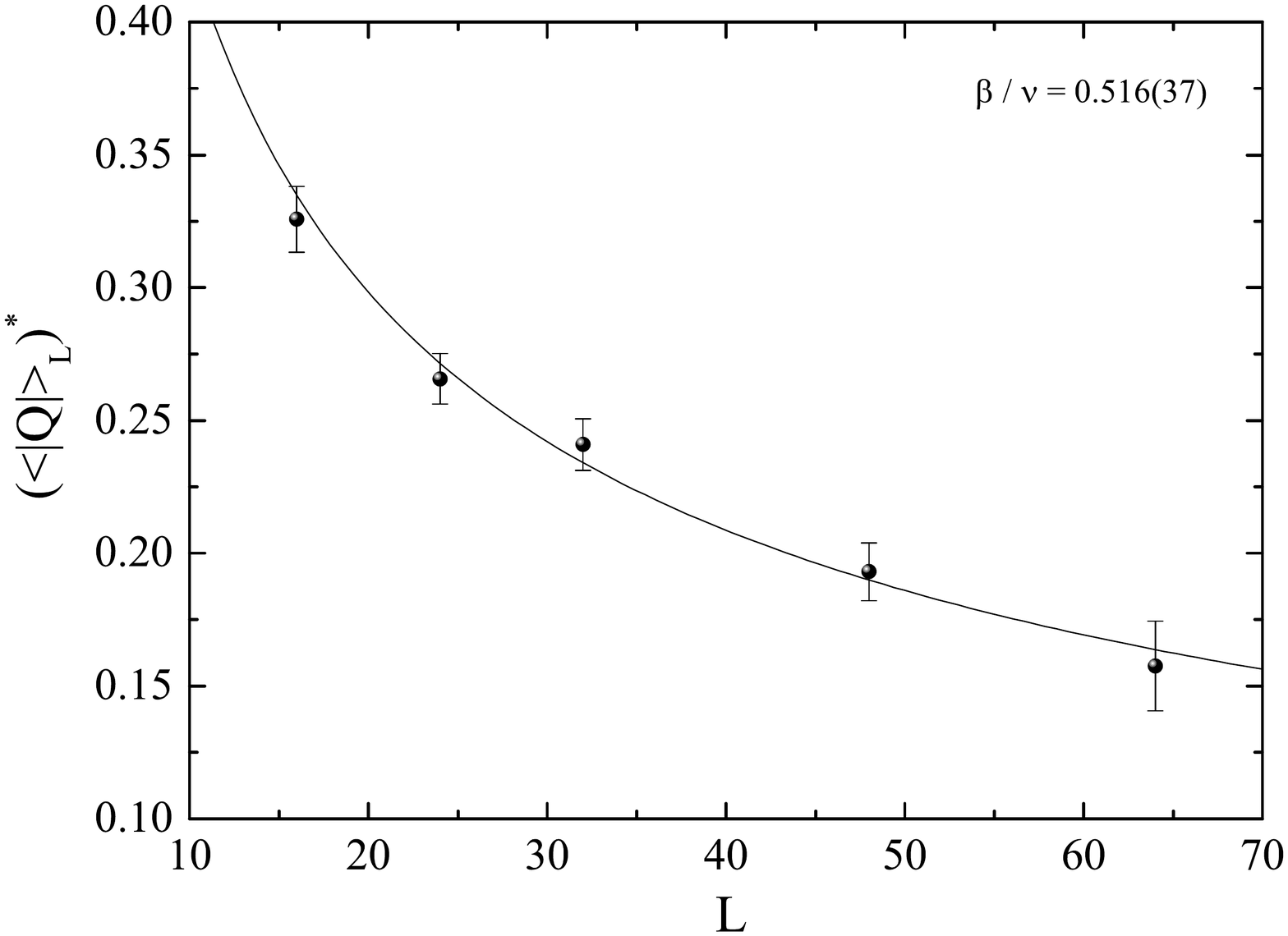}
\caption{\label{fig:14} Finite-size scaling analysis of the
dynamic order parameter estimated at the critical half period,
$(\langle |Q| \rangle_{L})^{\ast}$, of the 3D kinetic spin-1
Blume-Capel model. The solid line is a power-law fit of the
form~(\ref{eq:10}).}
\end{figure}

\begin{figure}[htbp]
\includegraphics*[width=12 cm]{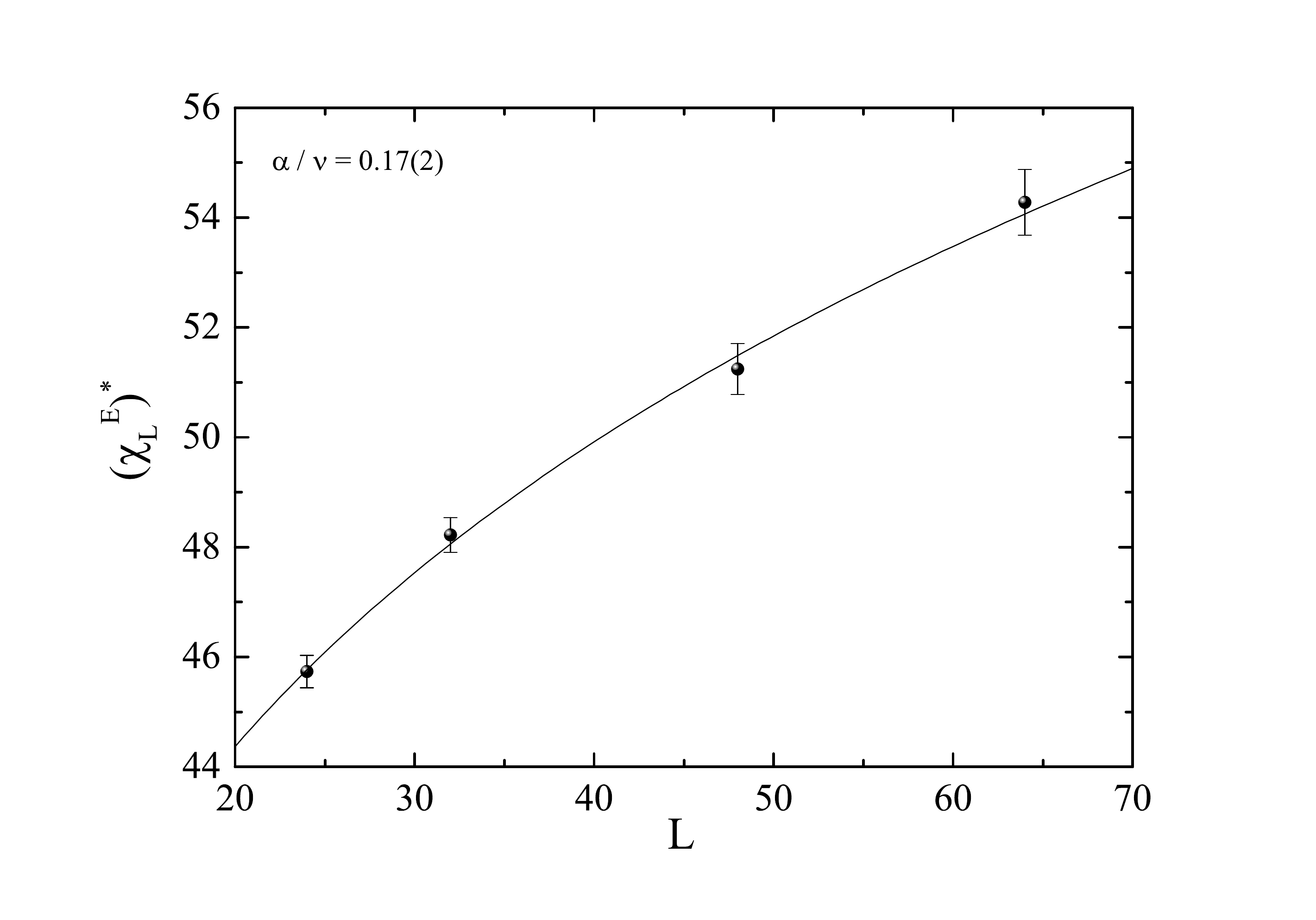}
\caption{\label{fig:15} Finite-size scaling analysis of the maxima
$(\chi_{L}^E)^{\ast}$ of the 3D kinetic spin-1 Blume-Capel model
The solid line is a power-law fit of the form~(\ref{eq:12}).}
\end{figure}


\begin{thebibliography}{100}
\bibitem{Tome} T. Tom\'{e} and M.J. de Oliveira, Phys. Rev. A \textbf{41},  4251 (1990).
\bibitem{Lo} W.S. Lo and R.A. Pelcovits, Phys. Rev. A \textbf{42},  7471 (1990).
\bibitem{Zimmer} M.F. Zimmer, Phys. Rev. E \textbf{47}, 3950 (1993).
\bibitem{Acharyya1} M. Acharyya and B.K. Chakrabarti, Phys. Rev. B \textbf{52}, 6550 (1995).
\bibitem{Chakrabarti} B.K. Chakrabarti and M. Acharyya, Rev. Mod. Phys. \textbf{71}, 847 (1999).
\bibitem{Acharyya2} M. Acharyya, Phys. Rev. E \textbf{56}, 1234 (1997).
\bibitem{Acharyya3} M. Acharyya, Phys. Rev. E \textbf{69}, 027105 (2004).
\bibitem{Buendia1} G.M. Buend\'{i}a and E. Machado, Phys. Rev. E \textbf{58}, 1260 (1998).
\bibitem{Buendia2} G.M. Buend\'{i}a and E. Machado, Phys. Rev. B \textbf{61}, 14686 (2000).
\bibitem{Jang1} H. Jang, M.J. Grimson, and C.K. Hall, Phys. Rev. E \textbf{68}, 046115 (2003).
\bibitem{Jang2} H. Jang, M.J. Grimson, and C.K. Hall, Phys. Rev. B \textbf{67}, 094411 (2003).
\bibitem{Shi} X. Shi, G. Wei, and L. Li, Phys. Lett. A \textbf{372}, 5922 (2008).
\bibitem{Punya} A. Punya, R. Yimnirun, P. Laoratanakul, and Y. Laosiritaworn, Physica B \textbf{405}, 3482 (2010).
\bibitem{Riego} P. Riego and A. Berger, Phys. Rev. E \textbf{91}, 062141 (2015).
\bibitem{Keskin1} M. Keskin, O. Canko, and U. Temizer, Phys. Rev. E \textbf{72}, 036125 (2005).
\bibitem{Keskin2} M. Keskin, O. Canko, and \"{U}. Temizer, J. Exp. Theor. Phys. \textbf{104}, 936 (2007).
\bibitem{Robb1} D.T. Robb, P.A. Rikvold, A. Berger, and M.A. Novotny, Phys. Rev. E \textbf{76}, 021124 (2007).
\bibitem{Deviren} B. Deviren and M. Keskin, J. Magn. Magn. Mater. \textbf{324}, 1051 (2012).
\bibitem{Yuksel1} Y. Y\"{u}ksel, E. Vatansever, and H. Polat, J. Phys.: Condens. Matter \textbf{24}, 436004 (2012).
\bibitem{Yuksel2} Y. Y\"{u}ksel, E. Vatansever, U. Akinci, and H. Polat, Phys. Rev. E \textbf{85}, 051123 (2012).
\bibitem{Vatansever1} E. Vatansever, Phys. Lett. A \textbf{381}, 1535 (2017).
\bibitem{He} Y.-L. He and G.-C. Wang, Phys. Rev. Lett. \textbf{70},  2336 (1993).
\bibitem{Robb} D.T. Robb, Y.H. Xu, O. Hellwig, J. McCord, A. Berger, M.A.
Novotny, and P.A. Rikvold, Phys. Rev. B \textbf{78},  134422
(2008).
\bibitem{Suen} J.-S. Suen and J.L. Erskine, Phys. Rev. Lett. \textbf{78},  3567 (1997).
\bibitem{Berger} A. Berger, O. Idigoras, and P. Vavassori, Phys. Rev. Lett. \textbf{111},  190602 (2013).
\bibitem{Riego1} P. Riego, P. Vavassori, and A. Berger, Phys. Rev. Lett. \textbf{118},  117202 (2017).
\bibitem{Sides1} S.W. Sides, P.A. Rikvold, and M.A. Novotny, Phys. Rev. Lett. \textbf{81}, 834 (1998).
\bibitem{Sides2} S.W. Sides, P.A. Rikvold, and M.A. Novotny, Phys. Rev. E \textbf{59}, 2710 (1999).
\bibitem{Korniss} G. Korniss, C.J. White, P.A. Rikvold, and M.A. Novotny, Phys. Rev. E \textbf{63}, 016120 (2000).
\bibitem{Buendia3} G.M. Buend\'{i}a and P.A. Rikvold, Phys. Rev. E \textbf{78}, 051108 (2008).
\bibitem{Park} H. Park and M. Pleimling, Phys. Rev. E \textbf{87}, 032145 (2013).
\bibitem{Tauscher} K. Tauscher and M. Pleimling, Phys. Rev. E \textbf{89}, 022121 (2014).
\bibitem{Vatansever2} E. Vatansever, arXiv:1706.03351.
\bibitem{Park2} H. Park and M. Pleimling Phys. Rev. Lett. \textbf{109}, 175703 (2012).
\bibitem{Buendia4} G.M. Buend\'{i}a and P.A. Rikvold, Phys. Rev. B \textbf{96}, 134306 (2017).
\bibitem{Capel} H.W. Capel, Physica (Amsterdam) \textbf{32}, 966 (1966).
\bibitem{Blume} M. Blume, Phys. Rev. \textbf{141}, 517 (1966).
\bibitem{Lawrie} I.D. Lawrie and S. Sarbach, in: C. Domb, J.L. Lebowitz (Eds.), \emph{Phase Transitions and Critical Phenomena}, Vol. 9 (Academic Press, London, 1984).
\bibitem{Selke} W. Selke and J. Oitmaa, J. Phys.: Condens. Matter \textbf{22}, 076004 (2010).
\bibitem{Berker1} A.N. Berker and M. Wortis, Phys. Rev. B \textbf{14}, 4946 (1976).
\bibitem{Branco} N.S. Branco and B.M. Boechat, Phys. Rev. B \textbf{56}, 11673 (1997).
\bibitem{Snowman} D.P. Snowman, Phys. Rev. E \textbf{79}, 041126 (2009).
\bibitem{Jain} A.K. Jain and D.P. Landau, Phys. Rev. B \textbf{22}, 445 (1980).
\bibitem{Falicov} A. Falicov and A.N. Berker, Phys. Rev. Lett. \textbf{74}, 426 (1995).
\bibitem{Silva} C.J. Silva, A.A. Caparica, and J.A. Plascak, Phys. Rev. E \textbf{73}, 036702 (2006).
\bibitem{Malakis1} A. Malakis, A.N. Berker, I.A. Hadjiagapiou, and N.G. Fytas, Phys. Rev. E \textbf{79}, 011125 (2009).
\bibitem{Malakis2} A. Malakis, A.N. Berker, I.A. Hadjiagapiou, N.G. Fytas, and T. Papakonstantinou, Phys. Rev. E \textbf{81}, 041113 (2010).
\bibitem{Malakis3} A. Malakis, A.N. Berker, N.G. Fytas, and T. Papakonstantinou, Phys. Rev. E \textbf{85}, 061106 (2012).
\bibitem{Fytas1} N.G. Fytas and W. Selke, Eur. Phys. J. B \textbf{86}, 365 (2013).
\bibitem{Kwak} W. Kwak, J. Jeong, J. Lee, and D.-H. Kim, Phys. Rev. E \textbf{92}, 022134 (2015).
\bibitem{Zierenberg} J. Zierenberg, N.G. Fytas, M. Weigel, W. Janke, and A. Malakis, Eur. Phys. J. Special Topics \textbf{226}, 789 (2017).
\bibitem{Boccara} N. Boccara, A. Elkenz, and M. Saber, J. Phys.: Condens. Matter \textbf{1}, 5721 (1989).
\bibitem{Hoston} W. Hoston and A.N. Berker, Phys. Rev. Lett. \textbf{67}, 1027 (1991).
\bibitem{Zahraouy} H. Ez-Zahraouy and A. Kassaou-Ou-Ali, Phys. Rev. B \textbf{69}, 064415 (2004).
\bibitem{ShiWei} X. Shi and G. Wei, Phys. Scr. \textbf{89}, 075805 (2014).
\bibitem{Acharyya4} M. Acharyya and A. Halder, J. Magn. Magn. Mater. \textbf{426}, 53 (2017).
\bibitem{Metropolis} N. Metropolis, A.W. Rosenbluth, M.N. Rosenbluth, A.H. Teller, and E. Teller, J. Chem. Phys. \textbf{21}, 1087 (1953).
\bibitem{Binder} D.P. Landau and K. Binder, \emph{A Guide to Monte Carlo Simulations
in Statistical Physics} (Cambridge University Press, Cambridge,
U.K., 2000).
\bibitem{Newman} M.E.J. Newman and G.T. Barkema, \emph{Monte Carlo Methods in Statistical Physics} (Oxford University Press, New York, 1999).
\bibitem{Glauber:63} R.J. Glauber, J. Math. Phys. {\bf 4} , 294 (1963).
\bibitem{Press}  W.H. Press, S.A. Teukolsky, W.T. Vetterling, and B.P. Flannery,
\emph{Numerical  Recipes  in  C},  2nd  ed.  (Cambridge University
Press, Cambridge, 1992).
\bibitem{Acharyya95} M. Acharyya and B.K. Chakrabarti, Phys. Rev. B \textbf{52}, 6550 (1995).
\bibitem{Binder81} K. Binder, Z. Phys. B: Condens. Matter \textbf{43}, 119 (1981); Phys. Rev. Lett. \textbf{47}, 693 (1981).
\bibitem{Fisher} M.E. Fisher, \emph{Critical Phenomena}, edited by M.S. Green (Academic, London, 1971).
\bibitem{Privman} V. Privman, \emph{Finite Size Scaling and Numerical Simulation of Statistical
Systems} (World Scientific, Singapore, 1990).
\bibitem{Binder92} K. Binder, \emph{Computational Methods in Field Theory}, edited by C.B. Lang and H. Gausterer (Springer, Berlin, 1992).
\bibitem{Ferrenberg} A.M. Ferrenberg and D.P. Landau, Phys. Rev. B {\bf 44}, 5081 (1991).
\bibitem{Ferdinand} A.E. Ferdinand and M.E. Fisher, Phys. Rev. \textbf{185}, 832 (1969).
\bibitem{Salas} J. Salas and A.D. Sokal, J. Stat. Phys. \textbf{98}, 551 (2000).
\bibitem{Selke_2} W. Selke, J. Stat. Mech. P04008 (2007).
\bibitem{Selke_3} W. Selke and L.N. Shchur, Phys. Rev. E \textbf{80}, 042104 (2009).
\bibitem{Hasenbusch} M. Hasenbusch, Phys. Rev. B \textbf{82}, 174434 (2010).
\bibitem{Kos} F. Kos, D. Poland, D. Simmons-Duffin, and A. Vichi, J. High Energy Phys. 08 (2016) 036.
\bibitem{Grinstein} G. Grinstein, C. Jayaprakash, and Y. He, Phys. Rev. Lett. \textbf{55}, 2527 (1985).
\end{thebibliography}
\end{document}